\documentclass{article}
\usepackage[utf8]{inputenc}
\usepackage[T1]{fontenc}
\usepackage{authblk}
\usepackage{amsmath, amssymb}
\usepackage{graphicx}
\usepackage{caption}
\usepackage{subcaption}
\usepackage{comment}
\usepackage{appendix}
\usepackage{soul}
\usepackage{hyperref}
\usepackage{geometry}
\usepackage{multirow}
\title{\textbf{Practical Use Cases of Neutral Atoms Quantum Computers}}

\author[1,2]{Matteo Grotti}
\author[3]{Sara Marzella}
\author[4]{Gabriella Bettonte}
\author[3]{Daniele Ottaviani}
\author[1,2]{Elisa Ercolessi}

\affil[1]{University of Bologna, Department of Physics and Astronomy, Bologna, Italy}
\affil[2]{Istituto Nazionale di Fisica Nucleare, Sezione di Bologna, Bologna, Italy}
\affil[3]{CINECA Quantum Lab, Bologna, Italy}
\affil[4]{E4 Computer Engineering, Scandiano, Italy}
\date{}

\begin{document}
\maketitle

\begin{abstract}
Quantum computing has quickly emerged as a revolutionary paradigm that holds the potential for greatly enhanced computational capability and algorithmic efficiency, in a wide range of areas. Among the various hardware platforms, neutral atom quantum processors based on Rydberg interactions are gaining increasing interest because of their scalability, qubit-connection flexibility, and intrinsic appropriateness for solving combinatorial optimization challenges. This paper provides an overview of the present capabilities, standards, and applications of neutral atom quantum computers. We first discuss recent hardware advancements and register mapping optimization techniques that enhance circuit fidelity and performance. We next review their uses as quantum simulators, in both classical and quantum hard problems, such as MIS and QUBO problems, quantum many-body models and molecules in chemistry and pharmacology. Applications for enhancing machine learning are also covered.
\end{abstract}

\vspace{0.5em}
\section{Introduction}

The literature on Quantum Computing and Simulations has surged in recent years, because these cutting-edge paradigms of computation are expected to bring disruptive improvements in our current computational capacity, enhancing the algorithmic performance in solving a huge variety of problems of interest for scientific and industrial applications. 
A variety of algorithmic methods have been proposed to tackle complex classical and quantum problems in both the digital \cite{Review_models_applications} and the analog approach \cite{Quantum_simulation_georgescu} \\
After the very first quantum computer was built using a Nuclear Magnetic Resonance (NMR) system \cite{NMR_QC}, different technologies for quantum hardware have been developed. These include platforms based on superconducting qubits \cite{Superconducting_qubits}, photonic circuits \cite{Photonics}, trapped ions \cite{trapped_ions}, NV centers \cite{NV_centers}, topological qubits \cite{Topological_QC}, and neutral atoms \cite{Pasqal,Quantum_computing_neutral_atoms,Quantum_computing_neutral_atoms_2,Aquila_Quera}. 
In the present Noisy Intermediate-Scale Quantum (NISQ) era \cite{QC_nisq_era_beyond}, no technology has yet emerged as better than the others, each of them having different advantages and limitations in terms of fidelity of the operations, scalability in the number of qubits, or the ability to perform error correction. 

In this review, we will focus on the neutral atom platform \cite{Pasqal,Quantum_computing_neutral_atoms,Quantum_computing_neutral_atoms_2,Aquila_Quera}, which has attracted a lot of attention because of some inherent features this technology can offer, such as the possibility to work at room temperature, long coherence time and high fidelity in operations of qubits, flexibility in qubit connectivity, and finally, the ability to implement error correction codes \cite{Logical_quantum_processor}. 
These hardware features make this technology interesting for the solution of a large class of problems hard to tackle on a classical computer.  For example, some classical optimization problems can be natively implemented on an atom-based platform, such as the Maximum Independent Set (MIS) problem\cite{MIS}, which is of fundamental importance for a wide variety of applications \cite{Industry_applications}. Also, this platform has been exploited for the simulation of quantum systems, such as many-body models \cite{Probing_many_body_dynamics_51_atoms} or chemical structures \cite{Analogue_quantum_chemistry_simulation}.
Finally, it has been adopted to enhance quantum machine learning capabilities \cite{QEK,Quantum_feature_maps_GML,Enhancing_GNN_QC_encodings,Financial_risk}, where typically used to optimize various machine learning models to solve classification or regression problems.

The scope of this paper is to summarize and review the wide range of applications for which neutral atom quantum computers have been considered. After providing a general overview of the problem, we will describe the principal advancements reached.  \\
The report is structured as follows. Sec. \ref{sec:QC} introduces the basic principles of quantum computing with neutral atom QPUs, whose performances are benchmarked in Sec. \ref{sec:benchmarks}.  Sec. \ref{sec:graph_problems} deals with classical optimization problems on graphs, focusing on the implementation of variational quantum algorithms to solve them. Sec. \ref{sec:physics} describes applications directly related to Physics, especially to quantum many-body theory, while applications to the fields of chemistry and pharmacology are presented in Sec. \ref{sec:Chemistry}. Sec. \ref{sec:machine_learning} is devoted to machine learning enhancement with the help of Rydberg atom arrays. Finally, the description of some other relevant problems, such as the SAT \cite{SAT_problem} and QUBO \cite{QUBO} ones, is provided in Sec. \ref{sec:other}. We draw some conclusions in Sec. \ref{se:conlusion}, providing some details on the hardware in the Appendix.The literature on Quantum Computing and Simulations has surged in recent years, because these cutting-edge paradigms of computation are expected to bring disruptive improvements in our current computational capacity, enhancing the algorithmic performance in solving a huge variety of problems of interest for scientific and industrial applications. 
A variety of algorithmic methods have been proposed to tackle complex classical and quantum problems in both the digital \cite{Review_models_applications} and the analog approach \cite{Quantum_simulation_georgescu}. \\
After the very first quantum computer was built using a Nuclear Magnetic Resonance (NMR) system \cite{NMR_QC}, different technologies for quantum hardware have been developed. These include platforms based on superconducting qubits \cite{Superconducting_qubits}, photonic circuits \cite{Photonics}, trapped ions \cite{trapped_ions}, NV centers \cite{NV_centers}, topological qubits \cite{Topological_QC}, and neutral atoms \cite{Pasqal,Quantum_computing_neutral_atoms,Quantum_computing_neutral_atoms_2,Aquila_Quera}. 
In the present Noisy Intermediate-Scale Quantum (NISQ) era \cite{QC_nisq_era_beyond}, no technology has yet emerged as better than the others, each of them having different advantages and limitations in terms of fidelity of the operations, scalability in the number of qubits, or the ability to perform error correction. 

Most of the application cases discussed in this review, such as QAOA and QAA-based optimization, quantum chemistry via VQE, and the simulation of lattice models, can be realized in principle on other quantum platforms, among which are superconducting qubits, trapped ions, photonics, and NMR systems. However, the practical feasibility of each of these implementations depends fundamentally on the native connectivity and coherence properties of the hardware, along with its control capabilities. In this respect, neutral-atom processors really offer singular advantages for problems whose natural form is a geometric or graph-constrained Hamiltonian, where the Rydberg interaction directly encodes the underlying combinatorial structure.

In this review, we will focus on the neutral atom platform \cite{Pasqal,Quantum_computing_neutral_atoms,Quantum_computing_neutral_atoms_2,Aquila_Quera}, which has attracted a lot of attention because of some inherent features this technology can offer, such as the possibility to work at room temperature, long coherence time and high fidelity in operations of qubits, flexibility in qubit connectivity, and finally, the ability to implement error correction codes \cite{Logical_quantum_processor}. 
These hardware features make this technology interesting for the solution of a large class of problems hard to tackle on a classical computer.  For example, some classical optimization problems can be natively implemented on an atom-based platform, such as the Maximum Independent Set (MIS) problem\cite{MIS}, which is of fundamental importance for a wide variety of applications \cite{Industry_applications}. Also, this platform has been exploited for the simulation of quantum systems, such as many-body models \cite{Probing_many_body_dynamics_51_atoms} or chemical structures \cite{Analogue_quantum_chemistry_simulation}.
Finally, it has been adopted to enhance quantum machine learning capabilities \cite{QEK,Quantum_feature_maps_GML,Enhancing_GNN_QC_encodings,Financial_risk}, where typically used to optimize various machine learning models to solve classification or regression problems.

The scope of this paper is to summarize and review the wide range of applications for which neutral atom quantum computers have been considered. After providing a general overview of the problem, we will describe the principal advancements reached.  \\
The report is structured as follows. Sec. \ref{sec:QC} introduces the basic principles of quantum computing with neutral atom QPUs, whose performances are benchmarked in Sec. \ref{sec:benchmarks}.  Sec. \ref{sec:graph_problems} deals with classical optimization problems on graphs, focusing on the implementation of variational quantum algorithms to solve them. Sec. \ref{sec:physics} describes applications directly related to Physics, especially to quantum many-body theory, while applications to the fields of chemistry and pharmacology are presented in Sec. \ref{sec:Chemistry}. Sec. \ref{sec:machine_learning} is devoted to machine learning enhancement with the help of Rydberg atom arrays. Finally, the description of some other relevant problems, such as the SAT \cite{SAT_problem} and QUBO \cite{QUBO} ones, is provided in Sec. \ref{sec:other}. We draw some conclusions in Sec. \ref{se:conlusion}, providing some details on the hardware in the Appendix. 

\section{Quantum Computing with neutral atoms}
\label{sec:QC}
\begin{figure}[t]
    \centering
    \subfloat[][\label{subfig:rydberg_levels}]{%
        \includegraphics[width=0.4\textwidth]{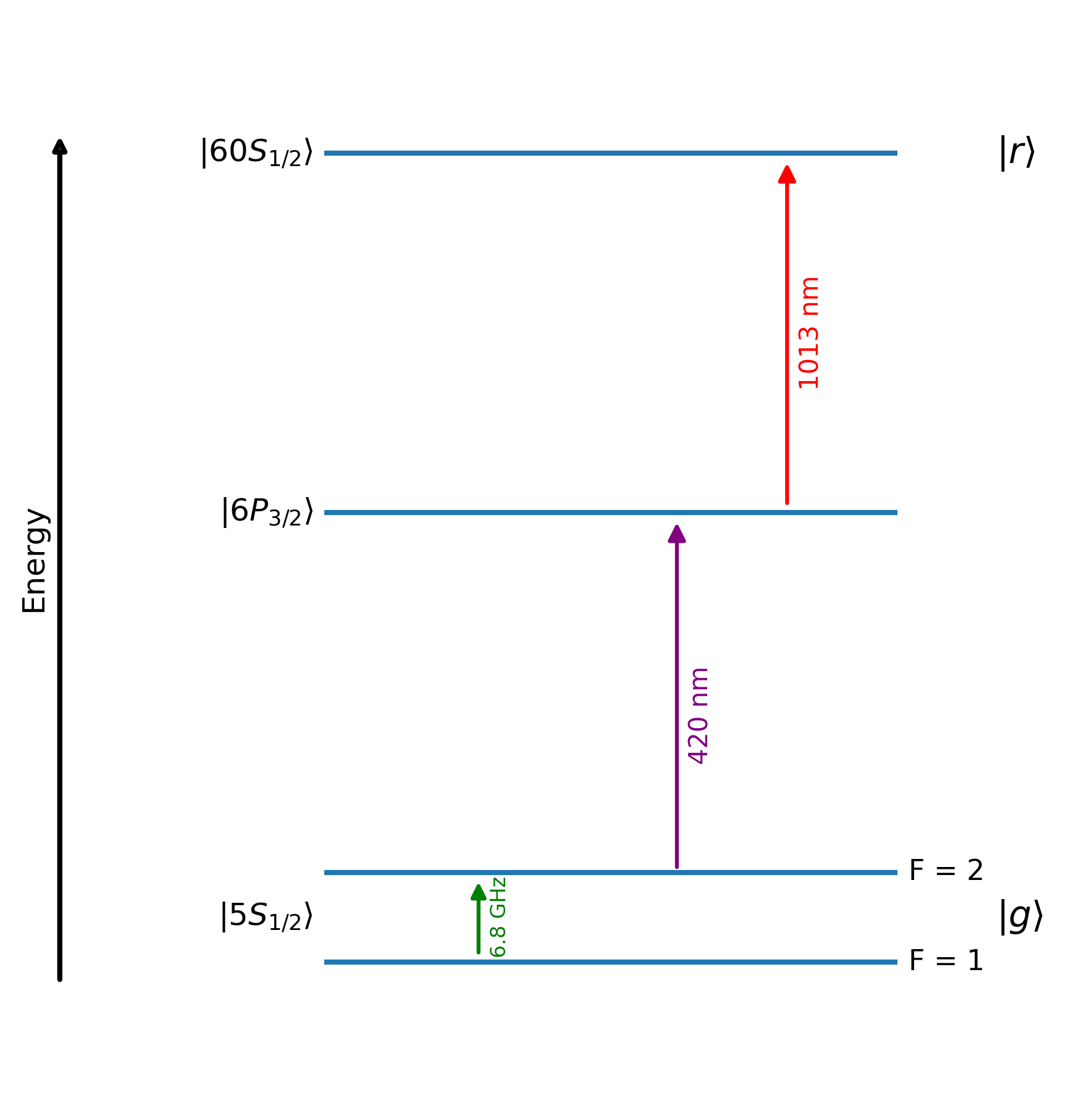}}
    \subfloat[][\label{subfig:rydberg_blockade}]{%
        \includegraphics[width=0.4\textwidth]{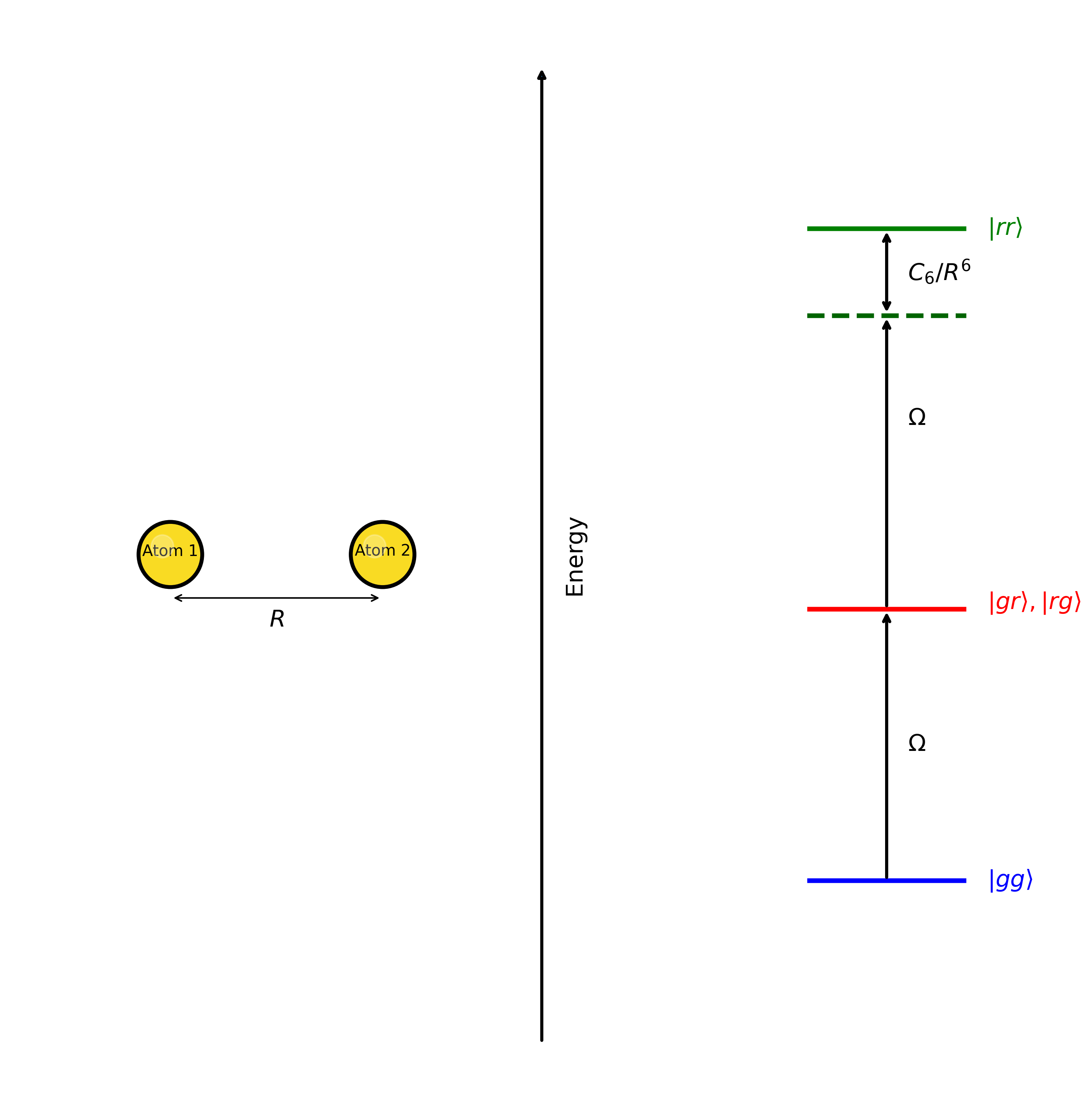}}
    \caption{(a) Atomic levels of Rubidium used for computation: the ground state $|5S_{1/2}\rangle$, the intermediate state $|6P_{3/2}\rangle$ and the Rydberg level $|nS_{1/2}\rangle$ where $50\leq n\leq 90$ (here we choose $n=60$). In the digital approach, the computational basis is formed by the hyperfine ground state with total angular momentum $F=1$ and $F=2$. In the analog approach the computational basis is given by the ground and the Rybderg level, and the hyperfine structure is neglected for the purpose of the computation. The atom can be excited to the Rydberg level via a two-photon transition scheme, including a first excitation to the intermediate level with a $420\; nm$ wavelength photon and a subsequent excitation to the Rydberg level via a $1013\;nm$ photon. (b) Rydberg blockade effect: when two atoms are at a distance $R$, the doubly excited level $|rr\rangle$ is shifted up in energy of a quantity $U=C_6/R^6$.}
    \label{fig:rydberg_levels}
\end{figure}
A neutral atom Quantum Processing Unit (QPU) works by trapping and cooling atoms, which typically belong to the alkali group (such as rubidium and cesium) or to the alkali-earth group (such as strontium) through optical tweezers \cite{Pasqal,Aquila_Quera}. The atoms, which encode the physical qubits are then placed in a graph structure (current technology allows to place atoms on a 2D lattice), called quantum register. The qubit basis states denoted as $|0\rangle$ and $|1\rangle$ can be encoded on different pairs of electronic states of the atom, depending on the usage mode, i.e. digital or analog. From now on, we will consider rubidium as it is the atomic species adopted by both QuEra and Pasqal companies, leading start-ups in the field of quantum computation with neutral atoms in USA and Europe, respectively. In digital mode, the basis states $\{|0\rangle,|1\rangle\}$ are encoded in the hyperfine structure levels of the electronic state $|5S_{1/2}\rangle$ with $F=1$ and $F=2$, respectively. In contrast, in the analog mode, the state $|0\rangle$ is encoded in the $|g\rangle=|5S_{1/2}\rangle$ ground state of rubidium, while the state $|1\rangle$ corresponds to the highly excited Rydberg level $|r\rangle=|nS_{1/2}\rangle$ with $50\leq n\leq 90$ \cite{Pasqal,Aquila_Quera}. Fig. \ref{subfig:rydberg_levels} shows a simple scheme of the electronic levels of Rubidium needed for computation. \newline The system dynamics is driven by means of laser pulses with a tunable Rabi frequency $\Omega$ (proportional to the laser amplitude), detuning frequency $\delta$ (i.e. the difference between the qubit resonance frequency and the actual laser frequency) and their relative phase $\phi$. The crucial key ingredient of quantum computing with neutral atoms are the strong Van der Waals interactions that are experimented by neighboring atoms in the excited Rydberg state \cite{Coincise_review_rydberg_atom_QCQI,QS_QC_rydberg_interacting_qubits,Rydberg_atom_quantum_technologies}, which hinder the simultaneous excitation of atoms whose distance is smaller than a threshold length known as the \textit{Rydberg blockade radius} \cite{Observation_rydberg_blockade}. This is due to a shift up in energy of the $|rr\rangle$ of one of the two atoms caused by the interaction term \cite{Coincise_review_rydberg_atom_QCQI} which physically decouples such state form the dynamics (see fig. \ref{subfig:rydberg_blockade}). As a consequence, neutral atoms represent a very powerful setup to achieve high-fidelity entanglement \cite{Towards_high_fidelity_QCQS_rydberg} since whenever a couple of qubits is placed within the blockade radius, the attempt to excite them would result in the final entangled state $\frac{|gr\rangle+|rg\rangle}{\sqrt{2}}$. Appendix from \ref{Appendix:QC_register} to \ref{Appendix:QC_partial} illustrate the details of digital and analog computational modes together with register loading and readout technicalities.  
Neutral-atom platforms also exhibit a number of unique capabilities that set them apart from other quantum technologies. First, programmable all-to-all connectivity through dynamical rearrangement of optical tweezers allows the physical geometry of the qubit register to be adapted to the structure of the problem. Second, Rydberg-mediated interactions, such as the blockade mechanism, naturally implement hard constraints  which maps directly onto the Hamiltonians of combinatorial problems. Third, the large scalability potential, arising from trapping hundreds of atoms in parallel, provide a versatile computational framework.
\section{Benchmarks and optimization of neutral atoms platforms}
\label{sec:benchmarks}
In this section we discuss how to improve the overall performance of neutral atoms platforms, first introducing register mapping optimization techniques, then illustrating typical benchmarks of neutral atoms platforms and finally describing some possible quantum error correction methods. We refer to Appendix \ref{Appendix:QC_noise} a characterization of the most relevant noise sources.
\subsection{Register mapping improvement}
\label{sec:benchmarks_algo}
\begin{figure}[t]
    \centering
    \includegraphics[width=0.7\linewidth]{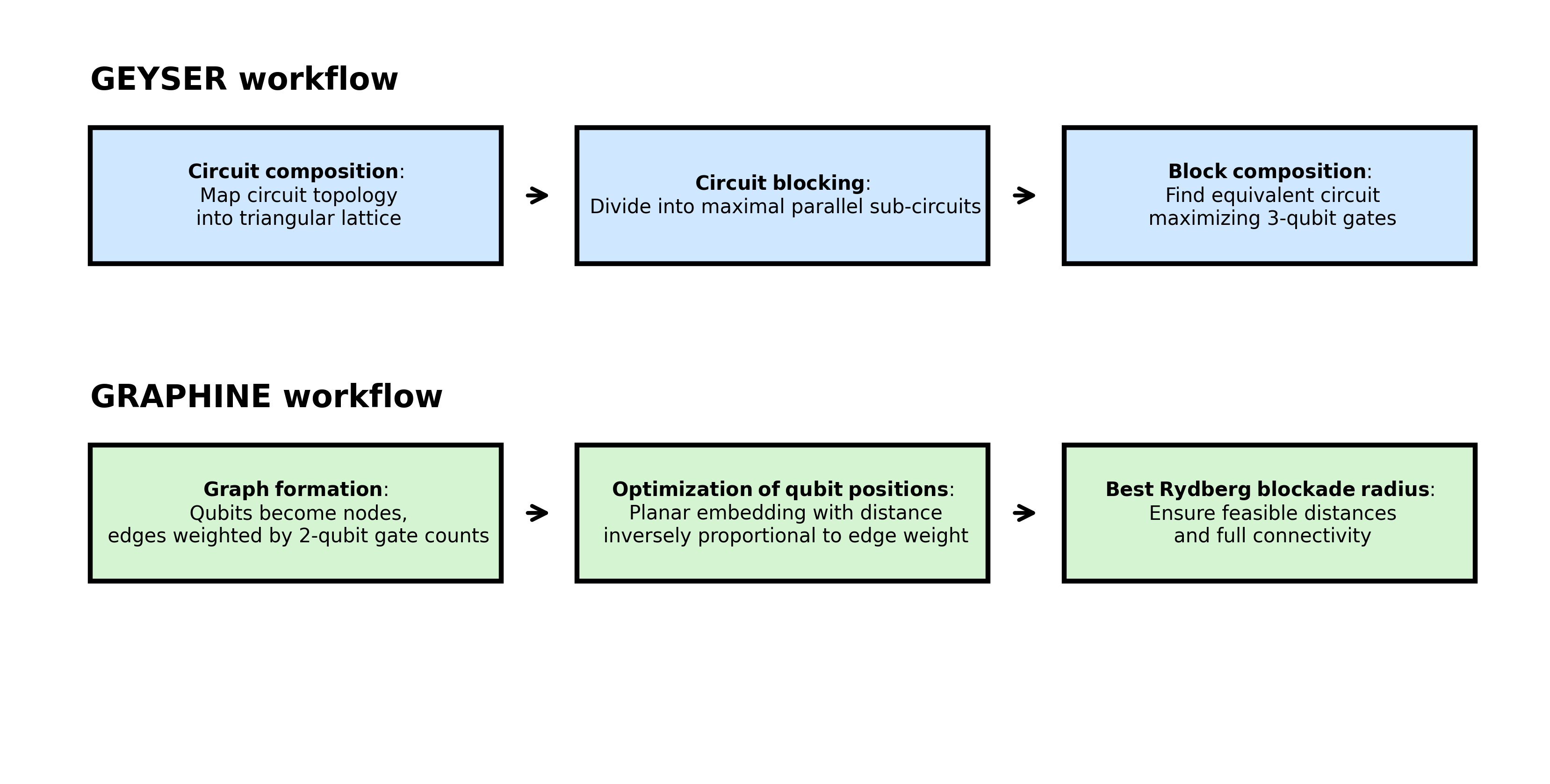}
    \caption{Workflow of Geyser \cite{Geyser} and Graphine \cite{Graphine}, whose details are described in the main text.}
    \label{fig:geyser_graphine}
\end{figure}
Since neutral atoms platforms allow one to place the qubits in arbitrary positions in the register, it is natural to wonder if there exist clever ways to select qubit connectivity based on the quantum circuit to be run and, in particular, based on the two qubit gates to be performed, as they require the atoms to be in the blockade regime. Tailored qubit connectivity allows for the optimal exploitation of the potentialities and advantages of neutral atoms devices compared to platforms with fixed topology such as superconduting qubits \cite{Superconducting_qubits}. Here we will discuss some possible answers to this fundamental question. The first approach taken into account is described in \cite{Optimal_mapping_rydberg}. It implements a swap gates insertion model \cite{Swap_gate_insertion_model} coupled with a one-dimensional topology displacement to minimize circuit depth, i.e. minimizing the inserted swap gates, and maximize circuit fidelity. This mapping technique has been benchmarked on arithmetic algorithms, random CNOTs generation, Bernstein Vazirani algorithm (BV) \cite{Bernstein-Vazirani_algo} and Quantum Fourier Transform (QFT) \cite{QFT, Optimal_mapping_rydberg}. Improvements on circuit depth of up to 70\% and of swap gate fidelity of up to 30\% have been achieved. The QFT benchmark shows a countertrend, as the model did not shorten circuit depth at all. \newline The next two mapping optimizers to be discussed are called GEYSER \cite{Geyser} and GRAPHINE \cite{Graphine}, devised in 2022 and 2023, respectively.\newline The workflow of GEYSER is the following:
\begin{itemize}
    \item Circuit composition: this first step aims at mapping the topology of the quantum circuit, meaning the necessary qubit connectivity needed to run the algorithm, into a lattice with triangular topology, typical of Pasqal's machines;
    \item Circuit blocking: the quantum circuit is blocked into sub-circuits which can be executed in parallel as possible;
    \item Block composition: the final step consists in finding, for each parallel block, an equivalent circuit which involves as many three-qubit gates as possible (hence reducing the total number of pulses) in order to speed up the computation.
\end{itemize}
The workflow of GRAPHINE is, instead, the following:
\begin{itemize}
    \item Graph formation: GRAPHINE creates a graph where qubits are nodes and the weights of the edges are based on the number of two-qubit gates between each qubit pair;
    \item Optimization of qubit position: the idea is to construct a planar lattice where the distance between each qubit pair is inversely proportioned to the corresponding edge weight;
    \item Find the best Rydberg blockade radius: the last step is to find the best blockade radius such that no two qubits are too distant and all the qubits are connected.
\end{itemize}
Fig. \ref{fig:geyser_graphine} schematically illustrates the workflow of both Geyser and Graphine. \newline
The main difference between GRAPHINE \cite{Graphine} and GEYSER \cite{Geyser} is that the former has no restriction on the lattice topology to be adopted while the latter embeds a circuit in a triangular lattice and this may limit, in principle, the quality of the embedding of GEYSER.
The performance of GEYSER in terms of number of gates, number of pulses and number of needed 1-, 2- and 3-qubit gates has been compared to the "baseline" (no mapping optimization) and a weakened version of GEYSER which does not perform circuit blocking and block composition on several quantum algorithms, including QFT, Quantum Approximate Optimization Algorithm \cite{farhi_qaoa} (QAOA) and Variational Quantum Eigensolver \cite{VQE_peruzzo} (VQE) with at most 16 qubits. Under all the parameters taken into account, GEYSER performs better than or equal to its weakened version and the baseline for all the circuit ansatz. \newline For what GRAPHINE is concerned, this protocol has been tested for the circuit optimization of several quantum algorithms including, in addition to the ones mentioned above, Hamiltonian evolution algorithms, specifically the XY \cite{XY_model} and Ising \cite{Ising_model} models, the Satisfiability \cite{SAT_problem} (SAT) problem solver and the adder and multiplier, which perform arithmetic operations. GRAPHINE has been compared to standard topology templates i.e. triangular \cite{Geyser}, squared \cite{Graphine_square_lattice} and hexagonal \cite{Graphine_hex_lattice} topologies. As a result, GRAPHINE performs better than or equal to its competitors for every ansatz. This is somehow expected since, as mentioned before, GRAPHINE has no restrictions on the lattice topology on which to run the algorithms. \newline The three algorithms presented here are tailored for the currently unavailable digital mode. However, the decrease of the circuit depth is crucial, since the coherence time of Rydberg atoms hinders the implementation of very deep circuits. 

\subsection{Benchmarking neutral atoms} 
\label{sec:benchmarks_benchmarks}

Benchmarking the performance of a quantum computer is not an easy task: there are many factors that have to be taken into account and that can influence the final outcomes in terms of the parameters with respect to which the benchmark is performed. Typical benchmarks are based on specific figures of merit devised to assess the performance of quantum platforms on specific quantum algorithms. This is the case of Q-Score \cite{Neutral_atom_hardware_performance}, a metric that is associated to the effective number of qubits that a specific quantum stack utilizes to solve the Maximum Cut problem;  the fidelity of the solution that can be  tested in the preparation of quantum states, such as the Bell pair $(|00\rangle+|11\rangle)/\sqrt{2}$ \cite{Quera_quantum_states,Multi-qubit_entanglement} or in the implementation of specific quantum algorithms, typically QAOA \cite{Multi-qubit_entanglement}, QFT, Deutsch-Josza \cite{Deutsch-Jozsa_algorithm,Deutsch-Jozsa_algo2} (DJ) or Bernstein-Vazirani \cite{Bernstein-Vazirani_algo} (BV) algorithms \cite{Benchmarking_neutral_atoms,Benchmarking_algorithmic_performance} together with Grover's one \cite{Grover_search_rydberg}. In the following, we present of summary of the main results, referring the interested reader to \cite{Nielsen_Chuang,Deutsch-Jozsa_algorithm,Bernstein-Vazirani_algo,QFT,Grover_algo} for the description of the standard DJ, BV, QFT and Grover algorithms, and to Appendix \ref{appendix:Algo} for a detailed discussion of the principal quantum optimization algorithms, namely the QAOA \cite{farhi_qaoa}, the Quantum Adiabatic Algorithm \cite{QAA} (QAA) and the VQE \cite{VQE}.

Ref. \cite{Benchmarking_highly_entangled_states_60-atom_rydberg} proves that analog neutral atoms devices are suitable to prepare highly entangled states in beyond-classical-tractability regime with high fidelity when $n\leq 60$ and compared them with MPS-based algorithm, showing that quantum computers can actually compete with state-of-the-art classical simulations. Wagner et al. \cite{Benchmarking_neutral_atoms} demonstrate that all-to-all connectivity of a Rydberg atoms QPU increases the fidelity (or success probability) of several algorithms included in the package developed by the QED-C (Quantum Economics Development Consortium), such as DJ, BV, hidden shift \cite{hidden_shift_algo} and QFT. The all-to-all connectivity has been compared with NN (nearest-neighbor) connectivity on a square grid, proving an advantage of the former in terms of fidelity and circuit depth. Notarnicola et al. \cite{Randomized_measurement_toolbox_rydberg} developed a randomized measurement toolbox based on a newly-devised protocol to implement arbitrary local rotations followed by measurements. This toolbox proves its efficacy in probing entanglement on different physical models, including the SSH \cite{SSH_model} and the XY \cite{XY_model} models. In 
\cite{Practical_quantum_advantage_QS}, taking into account possible sources of error for analog quantum computers \cite{Propagation_error_QS_quantum_advantage}, the authors proved a practical quantum advantage in comparison to current classical methods in the simulation of the dynamics of Ising model with transverse field \cite{Ising_model} and the Hubbard model \cite{hubbard_model_arovas} after a quench. For analog neutral atoms platforms, the current calibration level is sufficient to beat known classical algorithms, getting errors on the observables upper-bounded by 1\%.  

\subsection{Gates implementation and Quantum Error Correction} \label{sec:benchmarks_QEC}

In this subsection we will look at the possible strategies to increase two- or three-qubit gates fidelity. Subsequently, we will discuss some approaches concerning the emulation of digital gates. 
\begin{table}[t]
    \centering
    \begin{tabular}{|c|c|c|c|}
    \hline
    \multicolumn{2}{|c|}{CZ} &
    \multicolumn{2}{|c|}{CCZ} \\
    \hline 
        Input & Output & Input & Output \\
        \hline
        \multirow{2}{2em}{$|00\rangle$} & \multirow{2}{2em}{$|00\rangle$} & $|000\rangle$ & $|000\rangle$ \\
        & & $|001\rangle$ & $|001\rangle$ \\
        \multirow{2}{2em}{$|01\rangle$} & \multirow{2}{2em}{$|01\rangle$} & $|010\rangle$ & $|010\rangle$ \\
        & & $|011\rangle$ & $|011\rangle$ \\
        \multirow{2}{2em}{$|10\rangle$} & \multirow{2}{2em}{$|10\rangle$} & $|100\rangle$ & $|100\rangle$ \\
        & & $|101\rangle$ & $|101\rangle$ \\
        \multirow{2}{2em}{$|11\rangle$} & \multirow{2}{2em}{$-|11\rangle$} & $|110\rangle$ & $|110\rangle$ \\
        & & $|111\rangle$ & $-|111\rangle$ \\
        \hline
    \end{tabular}
    \vspace{1em}
    \caption{Effect of quantum gates CZ and CCZ on computational basis states.}
    \label{tab:CZ_CCZ}
\end{table}
\subsubsection{Gate fidelity improvement}
The future perspective for neutral atoms devices points towards the possibility to implement digital gates with local addressing laser pulses. This requires a considerable hardware improvement, consisting mainly in the massive development of the current capability to address single atoms with lasers. In addition, the correction of the main errors, which hinder a perfect implementation of quantum gates, has to be taken into account. Here we look at the strategies adopted to attain the highest possible fidelity for 2- and 3- qubit gates, typically the CZ (involved in the preparation of Bell pairs) and the CCZ, useful for more complex quantum algorithms. 
The effect of quantum gates CZ and CCZ on computational basis states are shown, for the sake of clarity, in Tab. \ref{tab:CZ_CCZ}.\newline
The fidelity of quantum gates has been kept increasing since the first attempts to realize multi-qubit gates on neutral atoms. Some of these attempts include \cite{Robust_quantum_logic_adiabatic} in 2015 where the Doppler-shifts noise caused by atomic motion has been tackled by devising a sequence with two counterpropagating lasers with opposite circular polarizations, achieving a CZ fidelity of 99.5\%. In \cite{High_fidelity_control_entanglement_rydberg}, a $|W\rangle$ entangled state has been prepared with a 97\% fidelity and, thanks to a dynamical decoupling protocol, which works by exciting the two atoms to the $|W\rangle$ state and de-exciting them back to $|gg\rangle$, its lifetime has been prolonged to $\sim 36 \mu s$. By parameterizing the adiabatic sequence to perform a CCZ gate, Tang et al. \cite{Fast_CCZ_gates_rydberg} devised an analog protocol reaching a 97.3\% gate fidelity. In \cite{High_fidelity_entanglement}, thanks to a single-mediated off-resonant modulated pulse (SORMD) pulse a 98\% fidelity Bell pair has been prepared. In subsequent works \cite{High_fidelity_gates_neutral_atoms}, this value has been further improved  to 99.5\%, reaching also a 97.9\% fidelity GHZ state preparation with the CCZ gate. Jandura et al. \cite{Optimizing_gates_logical_qubit} devised a new amplitude- and Doppler-robust pulse thanks to which the gate fidelity was 99.5\% or higher,  in the low-temperature regime in which the sequence is also independent from the temperature. Erasure errors are also tackled in \cite{Erasure_conversion_high_fidelity_rydberg_QS} for earth-alkali strontium atom: in this framework the authors achieve a Bell's state corrected fidelity of 99.85\%. 
QEC codes are an additional tool to achieve FTQC,. Some of these codes, like the well-known repetition code \cite{Repetition_code,Nielsen_Chuang}, the Steane code \cite{steane_code} and some surface codes like the XZZX \cite{XZZX_surface_code}, proved to be directly applicable on neutral atoms. We refer the interested reader to \cite{Hardware_efficient_QC_rydberg,Optimizing_gates_logical_qubit,Surface_code_connectivity} for more details. Last but not least, Machine Learning-based algorithm can be adopted to correct noise in quantum simulations. Mengoni et al. \cite{QML_noise_correction} devised a Reinforcement Learning-based protocol to adjust a simulated noise pulse, with gaussian-like Rabi frequency and linear sweep detuning, with the idea of compensating for the effects of the noise. The adjustments include modifications of the area of the $\Omega$-driven pulse as well as of the initial and final values of the detuning frequency. By looking at the trend of the Kullback-Leibler divergence \cite{KL_divergence} as a function of the number of epochs of the model (fig. 5 in \cite{QML_noise_correction}), they conclude the task was successfully achieved and their model was capable of correcting noise error. Artificial Neural Networks (ANN) \cite{Artificial_NN} have also been trained with the purpose of predicting the five noise parameters of a neutral atom device, namely laser intensity fluctuation, laser waist, temperature, false positive and false negative rates. In this regard, no good agreement between model prediction and real data was found. As authors suggest, this is may be due to the fact that real data, provided by Pasqal, have been taken while the machine was still under development. 

The fast improvement we observed in a few years toward almost-exact entangled state preparation and the hardware development highlight neutral atoms as a really promising candidate for robust gate implementation with long coherence times.

\subsubsection{Emulation of digital gates}
In this last subsection, we discuss a couple of approaches devised with the idea of emulating digital quantum computation with currently available analog simulators. 

Universal digital quantum computation can be achieved with only global laser pulses at the cost of a quadratic overhead in the number of atoms \cite{Universal_QC_global_rydberg_atoms}. This method works by building up a chain of qubits composed by two different atomic species in an alternated fashion alongside the chain. Additional impurities or superatoms (made out of in-blockade 4 atoms of the same species) are used to allow single qubit rotations. With a proper application of global unitaries on the two atom types, the state of the qubit (which is represented by an entire chain) flows forward on the chain and its state changes according to a single qubit gate, as shown in \cite{Universal_QC_global_rydberg_atoms}. The large number of atoms required to perform computations in this framework is the main drawback, which can be overcome only by scaling up the QPUs in the future. 

Another approach exploits partial addressing to emulate digital gates with non-local pulses \cite{Emulating_digital_gates}. By carefully choosing the evolution time under partial addressed pulses, local rotations around the X, Y and Z axis together with CZ two-qubit gate can be realized. Pulse optimization leads to fidelities up to 99.97\% for single qubit rotations performed on an interacting 1D chain with open boundary conditions and 99.7\% for the CZ gate on nearest-neighbors, which decreases to 94\% for the long-range SWAP (3 CNOTs) gate. The main drawback of this approach is the long time required to perform such gates (up to 18.7 $\mu s$ for the SWAP gate). Very deep circuits are beyond current capabilities because of the short coherence time compared to circuit duration ($6 \mu s$ have been achieved in \cite{QS_2D_antiferro_hundreds_qubits_rydberg}).
In general, possible hardware improvement directions in this sense could consist in increasing the laser power, in order to augment the maximum Rabi frequency and apply a larger number of quantum gates within the duration of the sequence, and decreasing the temperature at which the atoms are cooled down in order to reduce decoherence and prolong the sequence duration. 

\section{Graph problems}
\label{sec:graph_problems}
This section is dedicated to the discussion of graph problems, with a special focus on the Maximum Independent Set (MIS) \cite{MIS} and modern approaches to implement it and solve it on Rydberg atoms arrays. Section \ref{sec:MIS_def} provides the definition of three well-known graph problems, namely, the MIS, the MaxCut \cite{Maxcut} and the Max-Clique \cite{MaxClique}. In section \ref{sec:graph_problems_QAOAQAA} we discuss hybrid algorithms in order to tackle the MIS on neutral atoms, in conjunction also with quantum wires, described ultimately in section \ref{sec:graph_problems_wiring}. \newline Graph problems are of crucial importance due to their intrinsic potential to revolutionize our society in many research fields and the growing availability of advanced compilation frameworks for optimization on Rydberg hardware \cite{Quantum_toolkit_rydberg}. The development of quantum computers is expected to boost industrial processes \cite{Industry_applications}, chemistry and medicine, leveraging machine learning protocols \cite{GML_medicine,GML_chemistry}, with a special focus on drug discovery \cite{GML_drug_discovery}, molecular docking \cite{Molecular_docking_review}, and many body physics \cite{Probing_many_body_dynamics_51_atoms}.

\subsection{MIS, Max-Cut and Max-Clique}
\label{sec:MIS_def}
\begin{figure}[t]
    \centering
    \includegraphics[width=0.7\linewidth]{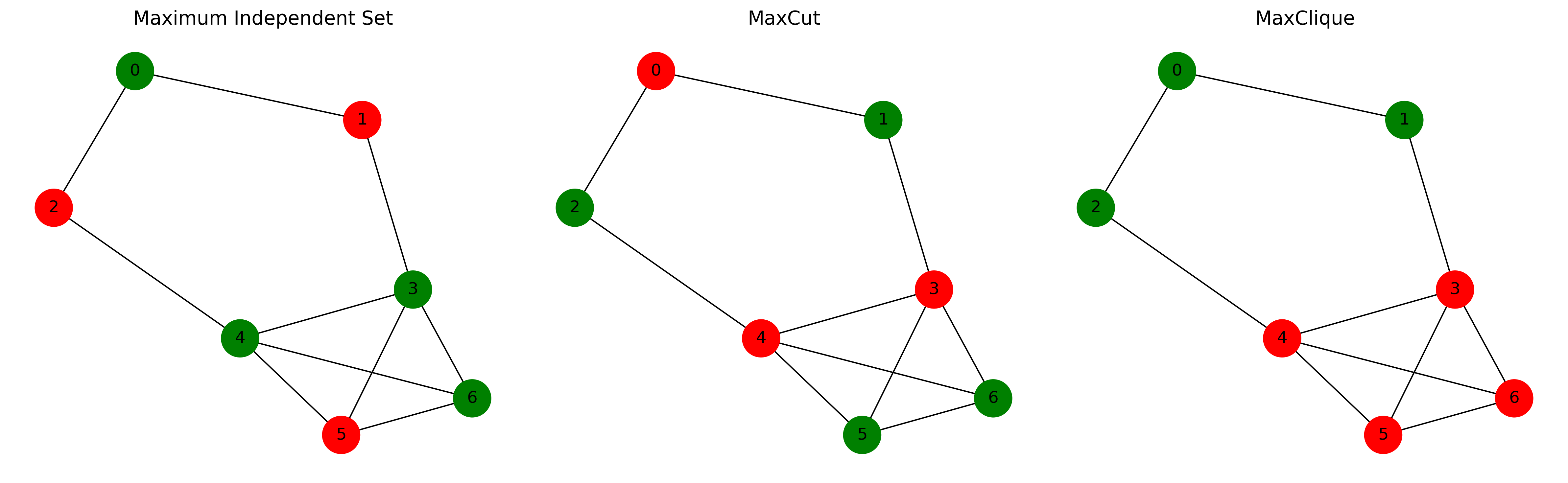}
    \caption{MIS, MaxCut and MaxClique solution for a simple graph instance}
    \label{fig:maxstuff}
\end{figure}
The Maximum Independent Set \cite{MIS} is a widespread, well-known graph problem. It is defined as follows: \newline \textit{Given a graph $\mathcal{G}=(V,E)$, with $V$ the set of vertices and $E$ the set of edges, the Maximum Independent Set asks to find a set of independent nodes, i.e. not connected to each other by any edge, of maximum cardinality}. \newline The solution of the MIS can be encoded into the ground state of the following Ising-like Hamiltonian: 
\begin{equation}
    H_{MIS}=\sum_i \frac{\sigma^z_i}{2}+\omega\sum_{\langle ij\rangle}\frac{\sigma^z_i\sigma^z_j-\sigma^z_i-\sigma^z_j}{4}.
    \label{eq:Hamiltonian_MIS}
\end{equation}
Notably, this Hamiltonian is equivalent to that in eq. (\ref{eq:Hamiltonian_pasqal}). As a consequence, the MIS is a native problem solvable on a neutral atom device since the ground state of the Hamiltonian (\ref{eq:Hamiltonian_pasqal}), which can be properly prepared, encodes the solution of the MIS. the more general Maximum Weight Independent Set (MWIS) problem, where edges in $E$ are given a weight, is more complicated and, from an hardware point of view, it can be handled by encoding the weights into the strength of the interactions between atoms, i.e. choosing a different distance between the atoms. Let us notice that not all possible graphs and sets of weight can be implemented in an exact way, as also highlighted in recent studies addressing hardware compatibility and embedding overhead on Rydberg arrays \cite{Quantum_toolkit_rydberg}, a problem that will be addressed in section \ref{sec:graph_problems_wiring}.

Other graph problems worth mentioning are: the MaxCut \cite{Maxcut}, defined as:
\newline \textit{Given a graph $G=(V,E)$, the MaxCut asks to find a bipartition of $V$ such that the number of edges connecting nodes belonging to different subsets is maximum} \newline and the MaxClique, given by \cite{MaxClique}: 
\newline \textit{Given a graph $G=(V,E)$, the MaxClique problem asks to find the subset of nodes of maximum cardinality such that the subgraph of $G$ made out only of those nodes is fully connected} \\
With some manipulations of the graph, both the MaxCut and the MaxClique can be mapped into a MIS \cite{MaxClique}. In particular, the MaxClique is equivalent to a MIS on the complementary graph of $G$. As a consequence, if a given problem can be written as either a MIS, a MaxCut or a MaxClique, it can be directly implemented and solved with neutral atom arrays. For completeness, we also mention that clustering aggregation problems \cite{clustering_as_mwis} can be restated in terms of a MIS.
Fig. \ref{fig:maxstuff} displays the solution for the three aforementioned graph problems on a simple 7-node instance. This graph has been properly chosen such that the solution node set is different for each problem.

To the best of our knowledge, the first proof of the possibility to solve the MIS on a neutral atoms device dates back to 2018 \cite{Quantum_opt_MIS}, working with a pretty straightforward implementation and the idea of improving available algorithms \cite{Combinatorial_graph_problems,Graph_Algo}. The developments pointed toward two main directions: the improvement of hybrid and variational algorithms and the development of techniques for mapping the graphs on the qubits register. Recent works have further advanced these directions by integrating quantum-informed reduction strategies \cite{qredumis} and by developing comprehensive compilation pipelines for mapping generic MIS instances onto Rydberg hardware \cite{Quantum_toolkit_rydberg}. This approaches will be discussed in sec. \ref{sec:graph_problems_QAOAQAA} and \ref{sec:graph_problems_wiring}, respectively.

\subsection{QAOA and QAA for MIS} \label{sec:graph_problems_QAOAQAA}

Although a standard version of QAOA is enough to find good approximate solutions, even in shallow circuit regimes and low repetition rate, especially if coupled to a bayesian optimizer \cite{Bayesian_QAOA}, a direct comparison against its main competitor, i.e. the Variational Quantum Adiabatic Algorithm (VQAA) which is an improved version of QAA where the adiabatic pulse is optimized classically, emphasizes how the QAOA can be outperformed by VQAA even at small effective depth \cite{Quantum_opt_MIS_rydberg_arrays}, at least for MIS instances up to 289 qubits. In particular, Ebadi et al. showed that hardware limitations, including leakage errors, blockade and pulse imperfections, hinder the classical optimizer to spot the best QAOA parameters and that's why, when increasing the circuit depth $p$, the approximation ratio reaches a maximum and then starts to fall again. The same happens for VQAA but the peak of the approximation ratio is much higher than for QAOA, reaching a maximum value greater than $90\%$ \cite{Quantum_opt_MIS_rydberg_arrays}. Complementary approaches have demonstrated that Bayesian Optimization can be used to design highly efficient quantum annealing schedules, achieving substantial improvements in fidelity for MIS instances on Rydberg devices \cite{Designing_quantum_annealing_bayesian}. Furthermore, VQAA has also been benchmarked with Simulated Annealing (SA) \cite{simulated_annealing}. The approximation ratio for SA at low depth is similar to that of VQAA. Nonetheless, for some graph instances with many local minima (corresponding to independent sets with one vertex less than the MIS) SA gets stuck and the approximation ratio displays a plateau until a very high depth is reached \cite{Quantum_opt_MIS_rydberg_arrays}). The performance of the VQAA is described by quasi-adiabatic evolution in agreement with the Landau-Zener formula \cite{Landau-Zener_formula}, which qualitatively indicates that the fidelity, i.e. the probability to measure a MIS solution bitstring, strongly depends on the energy gap between the ground state (MIS energy) and the first excited state: the higher the latter, the higher the fidelity \cite{Quantum_opt_MIS_rydberg_arrays}. 
Furthermore, recent analytical studies have shown that the time-to-solution for MIS instances on Rydberg devices is intrinsically constrained by graph-dependent hardness parameters and degeneracy structure, setting fundamental performance limits for adiabatic and annealing-based approaches \cite{Hardness_MIS}.

Some modified versions of QAOA have been presented over time. For the purposes of this review, we will consider FALQON \cite{Falqon_maxcut} and Non-Native Hybrid Algorithms (NNHA) \cite{NNHA_Quera}. \\
The former consists of a QAOA-like quantum evolution which does not require further classical optimization. The underlying idea is to achieve a monotonic decrease of the expectation value of the problem Hamiltonian over layers. This is done by computing the evolution parameter $\beta_j$ of the mixer layer $j$ from the expectation value of the commutator between the cost and the mixer Hamiltonians in the previous layer, i.e. $\beta_j=-k\langle\psi_{j-1}|i[H_M,H_C]|\psi_{j-1}\rangle$. By choosing $k=2$, which is enough to achieve $\frac{d}{dt}\langle H_C\rangle\leq 0$ according to the authors \cite{Falqon_maxcut}, a continuous decrease in energy $\langle H_C\rangle$ is induced. As a consequence, an approximate solution of the ground state naturally emerges at the end of the circuit. A benchmark of MaxCut graph instances with $n=2,3$ sites is provided, showing that the approximation ratio approaches $1$ as the number of layers increases. Unfortunately, larger system sizes are not taken into account. \newline The other aforementioned approach relies on NNHA \cite{NNHA_Quera}. The key ingredients of NNHA are a variational algorithm, which allows to sample bitstrings following a given distribution depending on the variational parameters and problem instance, and classical computation, which is entrusted with two main tasks: the optimization of variational parameters, as in standard hybrid algorithms, and the mapping of sampled bitstrings into valid problem solutions. This represents the real novelty brought by NNHA. The latter task can be fulfilled in several ways: post-processing of sampled bitstrings into individual solutions (e.g. using a Greedy flip \cite{Greedy_flip},  type-1), computation of correlation functions to infer system properties (type-2) and the usage of the sampled distribution as a reservoir (type-3). For each of these methods, \cite{NNHA_Quera} provides a use case. Type-1 post-processing is used to solve MaxCut with QAOA and Greedy flip, type-2 exemplification consists in the solution of Max-$k$-cut leveraging the low eigenvectors of the correlation matrix, while type-3 is employed to solve MIS with cluster-updated simulated annealing. Without delving into the details of each simulation, the overall performance of NNHA is comparable with that of classical-only algorithms.

\subsection{Quantum wiring and problem mappings} 
\label{sec:graph_problems_wiring}
\begin{figure}[t]
    \centering
    \includegraphics[width=0.7\linewidth]{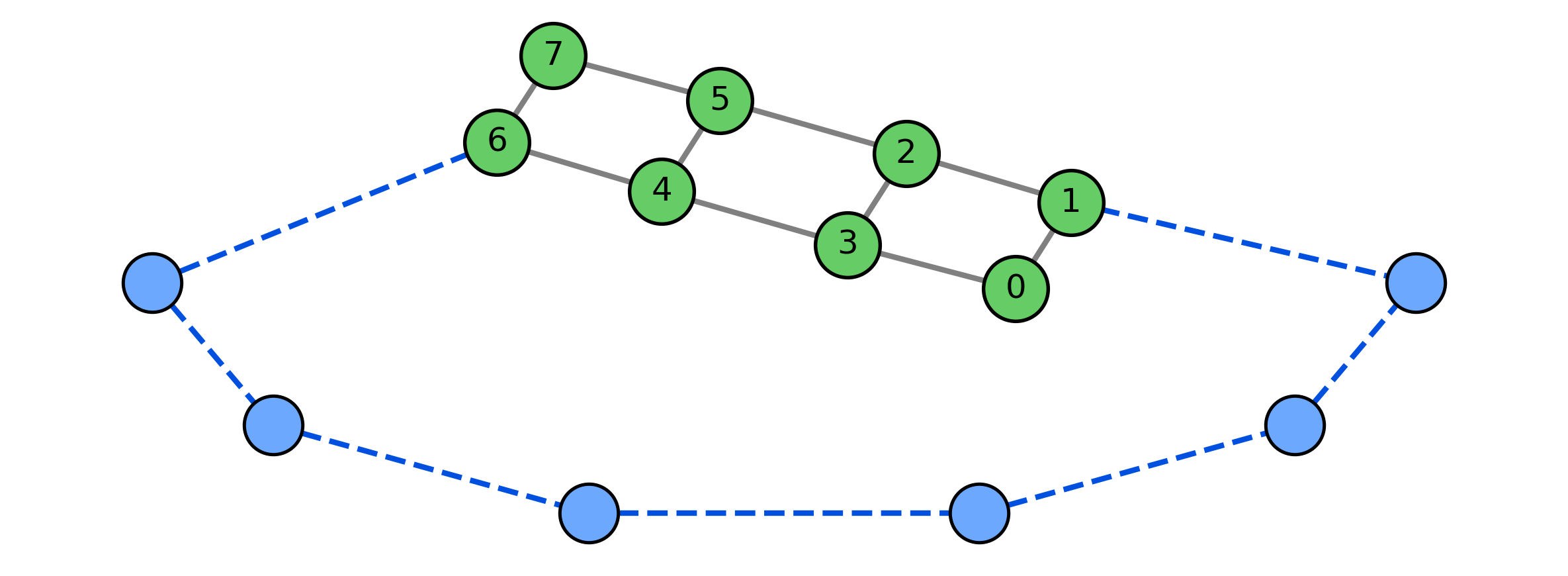}
    \caption{Example of quantum wiring on a 8-node atom arrangement. Nodes $1$ and $6$ are connected through a wire of an even number of atoms to ensure MIS constraint. Quantum wiring unlocks the possibility to solve the MIS problem on a non-UD graph by connecting far nodes with chains of extra atoms}
    \label{fig:quantum_wiring}
\end{figure}
By construction, the quantum register containing Rydberg atoms has limited connectivity, equivalent to that of a Unitary Disk (UD) graph. It is therefore fundamental to figure out how is it possible to address those graph problems which are natively non-UD. Quantum wires are of great help in this sense \cite{Rydberg_quantum_wires}. This approach consists in connecting far nodes with a chain of atoms of even length (to satisfy the MIS constraints), as shown in Fig. \ref{fig:quantum_wiring}. An alternative arrangement of the wires relies on orthogonal grid drawing (ODG) \cite{Orthogonal_grid_drawings} and it is presented in \cite{Graph_locality}. The main difference consists of the positions of the ancillary qubits, which are displaced in an orthogonal pattern \cite{Graph_locality}. The number of ancillary qubits scales sublinearly in the number of qubits of the original problem. An inverse mapping is provided in \cite{MIS_platonic_rydberg}, where the MIS has been solved on platonic solids mapped to 2D surfaces. 

Rydberg wires can be exploited to solve not only the MIS, but also QUBO \cite{QUBO} and SAT \cite{SAT_problem} problems. This is possible by mapping them into a MWIS, harnessing Rydberg gadgets as shown in \cite{Optimization_arbitrary_connectivity} by Nguyen et al. The drawback is a worse overhead in the number of qubits, which is quadratic in this case with respect to the OGD-based mapping. Nevertheless, this method entails a key advantage which consists in tackling a larger variety of problems. 
\section{QUBO and SAT problems}
\label{sec:other}
This section is devoted to illustrate how Quadratic Unconstrained Binary Optimization (QUBO) \cite{QUBO} and Satisfiability (SAT) \cite{SAT_problem} problems can be solved using neutral atoms devices. Further research is required, but the approaches discussed in sec. \ref{sec:other_QUBO} for the QUBO and in sec. \ref{sec:other_SAT} for the SAT pave the way toward future applications

\subsection{The Quadratic Unconstrained Binary Optimization problem}
\label{sec:other_QUBO}


The Quadratic Unconstrained Binary Optimization (QUBO) problem \cite{QUBO} consists in finding the bitstring $z^*=(z_1,...,z_n)$ such that the following quadratic cost function is minimized:
\begin{equation}
    f(z)=\sum_iQ_{ii}z_i+\sum_{i<j}Q_{ij}z_iz_j
    \label{eq:QUBO}
\end{equation}
Here, $Q_{ij}$ is called the QUBO matrix, whose size is $n\times n$, $n$ being the number of bits. The diagonal terms of $Q_{ij}$ define the linear term of the cost function while the quadratic term is given by the off-diagonal elements of the QUBO matrix. 

The similarity between the QUBO cost function (\ref{eq:QUBO}) and the Ising Hamiltonian (\ref{eq:Ising}) with a longitudinal field can be caught at a glance. This suggests that there should be a way to transform the QUBO problem into an Ising ground state preparation and, eventually, into a MWIS \cite{Tutorial_QUBO_models}. Concerning the resolution of a QUBO problem instance with neutral atoms devices, the main challenge to face consists in the embedding of the atoms in the quantum register. Roughly speaking, if one was able to find a qubit arrangement such that the interaction coefficient $U_{ij}$ between each pair of qubits $\{i,j\}$ matches the QUBO matrix entry $Q_{ij}$, the MWIS of the resulting register graph would be a solution to the QUBO. Needless to say, the exact embedding can be achieved only for a negligible minority of QUBO instances. In the general case, because of pure geometrical constraints, one has to settle for an non-ideal embedding which resembles the actual QUBO problem. 

An additional and more sophisticated alternative approach is offered by Byun et al. \cite{Rydberg_QUBO}. By splitting the graph in elementary local components and with the aid of Rydberg wires presented in sec. \ref{sec:graph_problems_wiring}, the authors propose a universal framework to encode QUBO on Rydberg atoms. The linear terms are encoded on data qubits: if $Q_{ii}<0$, the data qubits representing the same logical bit are not interacting with each other, such that the MIS of that subgraph correspond to qubits in the state $|1\rangle$; on the contrary, if $Q_{ii}>0$, ancillary qubits are connected to data qubits such that the local MIS can be obtained when the data qubits are in $|0\rangle$ while the ancillaries are excited . The quadratic terms are, instead, implemented by means of quantum wires, where the number of ancillary qubits forming each wire depends on the sign of $Q_{ij}$ (even for positive coefficient, odd for negative ones). The solution of the MWIS on the resulting graph is then found via adiabatic sweep. This method yields high fidelities. However, the scaling estimate of the number of qubits with respect to the size of the QUBO matrix is not provided. The main drawback of a Rydberg wires-based approach is that for more complex problem instances it requires a 3D register, although the instances presented in \cite{Rydberg_QUBO} can be implemented in 2D.

The QUBO model can be generalized to the Higher-order Constrained Binary Optimization problem (HCBO), whose Hamiltonian, in the spin formulation, reads:
\begin{equation}
    H=\sum_iJ_is_i+\sum_{ij}J_{ij}s_is_j+\sum_{ijk}J_{ijk}s_is_js_k+...
\end{equation}
where the variables $s_i=\pm 1$, $J_i$ are local fields and the coefficients $\{J_{ij},J_{ijk},...\}$ describe many-body interactions. Hamiltonians with interaction terms involving more than 2 qubits can be implemented via the LHZ parity encoding \cite{QAA_LHZ}. In this way, the higher order terms can be rewritten into parity constraints which can be subsequently mapped into a MWIS. In particular, the 3-body parity constraint becomes a 6-qubits MWIS to be solved. Several 3-body terms can be implemented in this way, and those building blocks can be connected via quantum wires \cite{Rydberg_blockade_parity}. 

Mixed Integer Linear Programming \cite{MILP} is a powerful tool to tackle optimization problems containing both discrete and continuous variables. In this context, Bender's decomposition \cite{BD} is of great help. It works by splitting the original problem (OP), whose variables can be either discrete or continuous, into a subproblem (SP), which includes the continuous variables of the OP, and a master problem (MP) containing the discrete variables. Naghmouchi et al. \cite{MILP_BD} addressed the MP with a neutral atoms quantum computer to speed up the computation. In particular, the MP can be reformulated into a QUBO and, ultimately, into a MWIS. QUBO embedding has been done with a heuristic algorithm devised for this purpose, while the solution of the MP has been found via the QAOA. In each step, the MP is rewritten as a QUBO and mapped into the atomic register. The QAOA solves the QUBO while the SP is solved classically. If the solution found does not obey the constraints of the MP, feasibility or optimality cuts are added to the QUBO and the process restarts. The algorithm ends when the solution does not violate any constraint. The proposed hybrid approach has been compared with a fully classical one, where the MP is solved by Simulated Annealing (SA). Both methods have been tested on 450 problem instances with at most 11 qubits, due to compatibility issues with real machines. It turns out that the QAOA outperforms SA in terms of percentage of feasible solutions found. It also maintains a closer distance to optimal solutions, classically computed beforehand \cite{MILP_BD}.  

\subsection{The Satisfiability problem}
\label{sec:other_SAT}

The Satisfiability problem (SAT) \cite{SAT_problem} is a boolean problem asking to determine whether a logical proposition, depending on boolean variables, is satisfiable or not. In case it is, the boolean variable string is to be found. The formulation of the SAT problem is typically provided in conjunctive normal form, given by the conjunction of $N_C$ clauses, where each clause is formed by the disjunction of elementary literals. Here we will focus on the 3-SAT problem, where the literals are at most 3. The 3-SAT proposition can be written as \cite{SAT_rydberg}:
\begin{align}
    &\Psi(x_1,...,x_n)=\bigwedge_j^{N_C} C_j \\
    &C_j=l_{j,1}\vee l_{j,2}\vee l_{j,3} \nonumber \\
    &l_{i,j}\in\{x_k,\bar{x_k}|k=1,...,n\}. \nonumber 
\end{align}

The 3-SAT problem can be reduced to the following MIS on a graph $G(V,E)$ \cite{SAT_rydberg}, with:
\begin{align}
    &V=\{(j,k)|l_{j,k}\in C_j\} \\
    &E=E_1\cup E_2 \nonumber \\
    &E_1=\{[(j,k_1),(j,k_2)]|k_1\neq k_2\} \nonumber \\
    &E_2=\{[(j_1,k_1),(j_2,k_2)]|k_1\neq k_2,l_{j_1,k_1}=\bar{l}_{j_2,k_2}\}. \nonumber
\end{align}
To each literal, a qubit on the register is assigned. Effectively, each clause $C_j$ is satisfied iff only one literal is true, i.e. the qubit corresponding to that literal is in the state $|1\rangle$ while the others are in their ground state. Therefore, a clause can be implemented on a Rydberg array by arranging the qubits within the blockade radius of each other. Van der Waals interactions will hence allow the excitation of only one of them. If a literal/qubit is involved in more than one clause, chains of ancillary qubits, i.e. odd-length Rydberg wires, can be exploited to reproduce the same qubit in two register positions far away from each other \cite{SAT_rydberg}. Jeong et al. \cite{SAT_rydberg} tested this framework on 3 3-SAT instances on a neutral atoms platform, solving the MIS with the adiabatic protocol. It has been found that the most probable bitstrings were actual solutions to the 3-SAT. Furthermore, it was estimated that the scaling of the number of atoms $n$ required to implement the MIS with respect to the number of clauses $N_C$ of the associated 3-SAT is upper bounded by $n\sim 4.88N_C^{1.8}$ and lower bounded by a linear function.  

\subsubsection{Integer Factorization}

The integer factorization problem consists in finding the prime factors of a given number. One of its main application is cybersecurity \cite{Integer_factorization_cybersecurity}, where many cryptography systems rely on the factorization of very large numbers. Integer factorization can be reduced to a SAT and, as we discussed in the previous section, to a MIS. With an approach similar to \cite{SAT_rydberg}, Park et al. \cite{Integer_factorization} showed how to implement and solve the factorization problem for 3 semiprime instances, namely $6=3\times 2$, $15=5\times 3$ and $35=7\times 5$. By means of QAA followed by postprocessing, the latter consisting in discarding bitstring which do not satisfy all the clauses, the prime factors of the above numbers have been found with very high fidelity. Nevertheless, the factorization of the number $15$ required the employment of a 3D register, because of geometrical reasons \cite{Integer_factorization}.
\section{Physics applications} \label{sec:physics}
In this section we will focus on the emerging applications of quantum simulations with Rydberg atoms arrays in physics \cite{Review_QS_rydberg}. We will start by discussing the implementation of the renowned Ising and Heisenberg models (sec. \ref{sec:physics_ising}), proceeding then to spotlight applications to fundamental interactions (sec. \ref{sec:physics_spin_liquids}).

\subsection{Ising and Heisenberg models} \label{sec:physics_ising}
Since Rydberg atoms has been deemed a valuable and promising platform for quantum computing purposes because of the dynamics of Rydberg atoms \cite{QS_energy_transport_embedded_rydberg_aggregates,long_range_interaction_rydberg,Probing_many_body_dynamics_51_atoms,Detailed_balance_rydberg,Quantum_kibble_zurek_mechanism,Many-body-physics_rydberg,Coherent_dynamics_rydberg_atoms,Controlling_quantum_many_body_dynamics_rydberg,Landau-forbidden_quantum_criticality_rydberg,Benchmarking_QS_ergodic_dynamics,Quantum_dynamics_fully_blockaded_rydberg_atom_ensemble}, many efforts have been aimed at exploring the capability of this architecture to implement the Ising model \cite{Ising_model} and the Heisenberg model \cite{Heisenberg_model,XXZ_model}, to find their ground states and possibly reproduce their unique magnetic and dynamical properties. This is because the Hamiltonians of these models:
\begin{equation}
    H_{ISING}=-\sum_{\langle ij\rangle}J_{ij}\sigma^z_i\sigma^z_j-h_x\sum_i\sigma^x_i-h_z\sum_i\sigma^z_i; 
    \label{eq:Ising}
\end{equation}
\begin{equation} 
    H_{HEI}=-\sum_{\langle ij\rangle}(J_x\sigma^x_i\sigma^x_j+J_y\sigma^y_i\sigma^y_j+J_z\sigma^z_i\sigma^z_j)-h_z\sum_i\sigma^z_i.
    \label{eq:Heisenberg}
\end{equation}
look incredibly similar to the ones driving the evolution of neutral atoms in the Ground-Rydberg and $XY $-basis (eq. (\ref{eq:Hamiltonian_pasqal}) and (\ref{eq:Pasqal_XY}), respectively). Here $\sigma_x,\sigma_y, \sigma_z$ are standard Pauli operators. We will delve into the practical realizations of these models on neutral atoms devices in the next subsections. 

\subsubsection{Ising model ground state in the Ground-Rydberg basis}
\begin{figure}[t]
    \centering
    \includegraphics[width=0.7\linewidth]{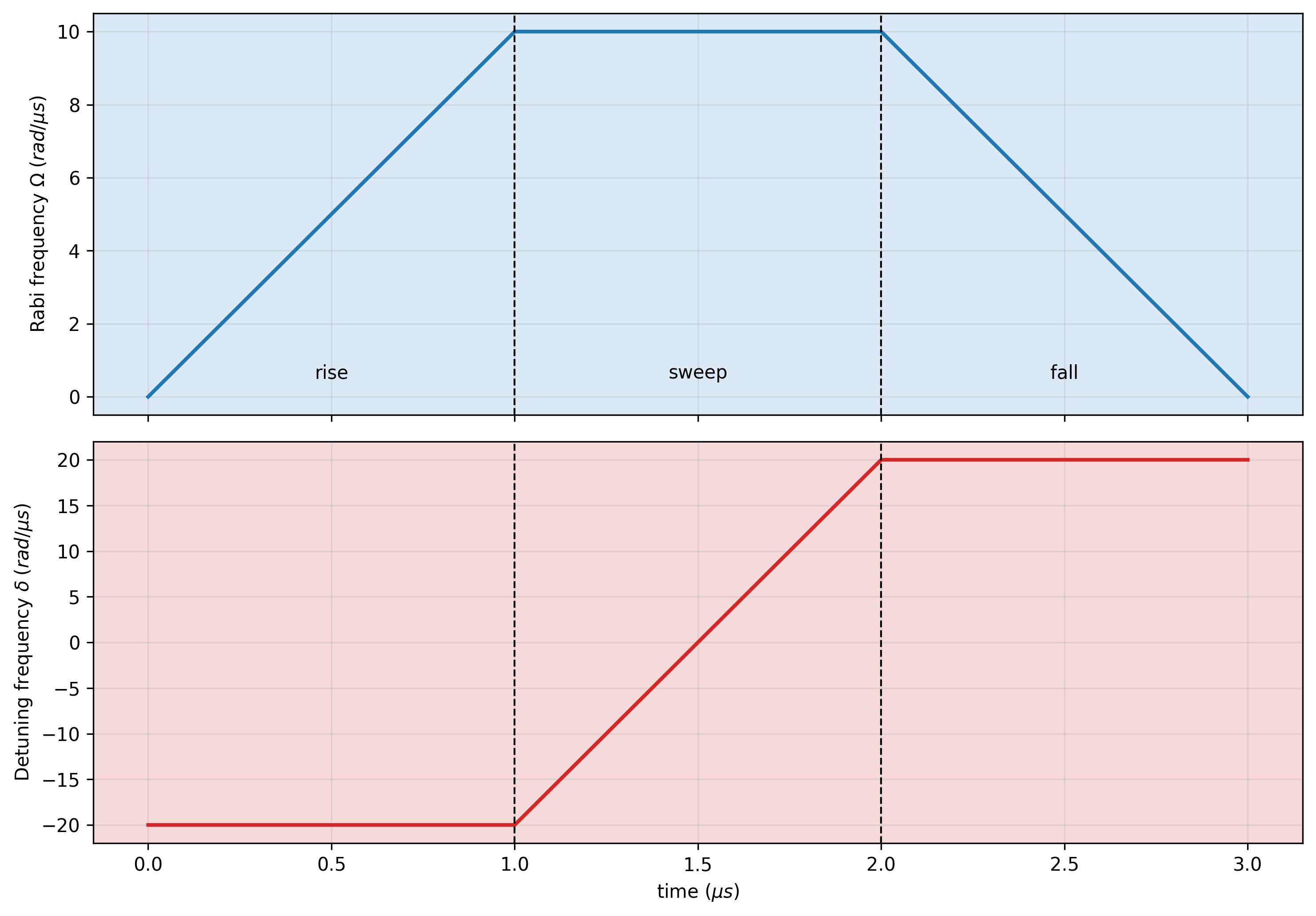}
    \caption{Adiabatic sequence to prepare the ground state of the antiferromagnetic Ising model. The algorithm consists of three main phases called "rise", "sweep" and "fall": starting from a far negative detuning $\delta$ and a null Rabi frequency $\Omega$, the latter is raised up to a maximum value $\Omega_{max}$; after that, while keeping $\Omega=\Omega_{max}$, the detuning is brought to a far positive value $\delta_{max}$ and, at the very end, the Rabi frequency is turned off.}
    \label{fig:adiabatic_seq}
\end{figure}
Far from being a trivial and effortless task, the preparation of the ground state of the Ising model with high fidelity, with a focus on the antiferromagnetic ($J=-1$) version, has attracted a lot of attention since the dawn of quantum computation with Rydberg atoms \cite{Realizing_quantum_ising_models_rydberg,Tunable_2D_array_quantum_ising,Simulating_quantum_spin_models_rydberg_magnetic_microtraps,Multilevel_rydberg_Ising,transverse_ising_rydberg}. A standard annealing scheme, in the Ground-Rydberg encoding basis, is typically employed to accomplish this goal. The full sequence is provided in Fig. \ref{fig:adiabatic_seq}. Basically, it works as follows:
\begin{itemize}
    \item The atoms are initialized in the state $|000...\rangle$;
    \item The Rabi frequency $\Omega$ is raised from 0 to a maximum value $\Omega_{max}$ while the detuning $\delta$ is kept constant to a (typically negative) initial value $\delta_i$ for a time $t_{rise}$;
    \item For a time $t_{sweep}$, $\Omega$ is left constant while $\delta$ is increased to a (typically positive) value $\delta_f$;
    \item In the final phase, which lasts a time $t_{fall}$, $\Omega$ is brought back to 0 with constant detuning;
    \item A measurement is performed as final step.
\end{itemize}
Improved or modified versions of this algorithm can be adopted depending on the specific parameters of the model. With the usage of partial addressing to drive local detuning on edge atoms, a 20-atom 1D GHZ state preparation with fidelity larger than  50\% has been achieved in \cite{Generation_schrodinger_cat_state_rydberg}. A few years later, Scholl et al. \cite{QS_2D_antiferro_hundreds_qubits_rydberg} increased the system size by an order of magnitude, implementing a 2D Ising model with at most 196 qubits, although reaching a quite low fidelity of the order of $1\%$, examining its magnetic properties both on a square and a triangular lattice. Meanwhile, Ebadi et al. \cite{Quantum_phases_of_matter_256_atom_QS} achieved an analogous result on a 256-qubits processor. In the same year, Ising model on Cayley-tree graphs has been tackled \cite{Cayley_tree_rydberg}, with analogous outcomes. 

An additional topic of investigation is represented by amorphous magnets \cite{amorphous_magnets_book}, which can be modeled by means of an Ising model \cite{Amorphous_quantum_magnets}, hence implementable on neutral atoms. Rydberg platforms have been deemed promising to characterize the collective properties of atoms that do not feature long-range order, as classical simulations are computationally expensive \cite{Amorphous_quantum_magnets}.

\subsubsection{Heisenberg model in the XY basis}

The XY basis, consisting of two Rydberg states with different angular quantum number $l$ experiencing dipole-dipole interactions, is suitable to evolve the system according to the Hamiltonian (\ref{eq:Pasqal_XY}). In order to widen the range of models for quantum simulation purposes, a sophisticated approach, called Floquet cycle \cite{Floquet_engineering_many_body}, has been devised to rotate the interaction axis of the Rydberg atoms  \cite{Microwave_engineering_XXZ} , leading the system to evolve under various Hamiltonians, including the XXZ Heisenberg one (eq. (\ref{eq:Heisenberg})). More specifically, by applying four global pulses with different phases ($0,-\pi/2,\pi/2,\pi$), on average the system will evolve according to:
\begin{equation}
    H_{av}=\sum_{ij}\frac{J_{ij}}{t_c}\biggl[(\tau_1+\tau_2)\sigma^x_i\sigma^x_j+(\tau_1+\tau_3)\sigma^y_i\sigma^y_j+(\tau_2+\tau_3)\sigma^z_i\sigma^z_j\biggr].
    \label{eq:Floquet_Hamiltonian}
\end{equation}
By properly tuning the waiting times $\tau_i$, one can engineer not only the XXZ Hamiltonian but also the general XYZ and Dzyaloshinskii-Moriya (DM) Hamiltonians \cite{QS_spin_exchange_models_floquet_rydberg} as well as dimerized models \cite{Thouless_pumps}. Despite the Floquet cycle represents an efficient protocol, Kim et al. \cite{Realization_anisotropic_heisenberg_magnet_rydberg} realize a Heisenberg magnet in an alternative way which leverages Van der Waals interactions (i.e. Ground-Rydberg basis) and explore the dynamics of single-magnon and two-magnon states.  Finally, we remark that the XY basis is also useful to investigate ring current states, i.e. superposition of states with delocalized Rydberg excitations in atomic rings \cite{Quantum_superpositions_rydberg_networks}.

\subsection{Fundamental interactions}
\label{sec:physics_spin_liquids}

This short section is devoted to applications to fundamental interactions in physics. This is a very wide topic, that would deserve a whole paper in its own. Here we just mention results about some prototypical examples related to topological matter and particle physics lattice gauge theories.

Topological Mott insulators \cite{Electronic_mott_insulator,Mott_insulator_2} are realized with a Hamiltonian consisting of a fermionic hopping term together with a staggered potential, with the addition of nearest-neighbor and next-nearest-neighbor interactions, Dauphin et al. analyzed how it is possible to recover the Quantum Anomalous Hall phase (QAH) \cite{Quantum_anomalous_hall_effect} through Rybderg atoms placed at the vertices and edges of a square (Lieb) lattice, properly tuning the laser parameters controlling atomic transitions \cite{QS_topological_mott_insulator_rydberg_lieb_lattice}. 
More generically, fermionic dynamics has been investigated with neutral atom set-ups \cite{Fermionic_quantum_processing}. For example, in \cite{Hubbard_physics_rydberg_spin_simulator_strong_fermionic_correlations}, a hybrid quantum-classical approach, based on slave-spin techniques \cite{Spin_slave_rep}, is used to simulate the Hubbard model
\cite{hubbard_model_arovas} on the square lattice in and out of equilibrium, using a Rydberg-based analog processor to solve the resulting interacting spin problem. 
Other intriguing models that can be naturally simulated on Rydberg platforms are Quantum Spin Liquids (QSL), a highly entangled, locally disordered phase of matter which exhibits long-range order and anyonic excitations \cite{Quantum_spin_liquids}. This can be obtained via a standard adiabatic protocol, by placing the atoms on a Kagome lattice and properly tuning the Rydberg blockade radius to implement dimer constraints \cite{Quantum_dimer_models_review,Quantum_phase_rydberg_kagome_lattice,Probing_topological_spin_liquids}. The ground state phase properties of this system have been compared with the one computed by time-dependent Density Matrix Renormalization Group techniques \cite{DMRG}, which classically emulates the adiabatic state preparation, finding good agreement between them \cite{Quantum_phase_rydberg_kagome_lattice,Probing_topological_spin_liquids,Dynamical_prep_quantum_spin_liquids_rydberg}.
Zeng et al.\cite{Dimer_models} proved that it is possible to prepare QSL topological states by adopting Rydberg gadgets, i.e. given arrangements of Rydberg atoms designed to yield specific constraints thanks to lattice geometry and the blockade effect. This approach has been applied to recover dimer models on different lattice topologies. More specifically, it has been argued that the preparation of Resonating Valence Bond (RVB) states is feasible on square and triangular lattices with a fidelity up to 99\%, comparable to the one obtained in \cite{Dynamical_prep_quantum_spin_liquids_rydberg}. 

Much work has been done to show how lattice gauge theories \cite{lattice_gauge_theory} naturally emerge from Rydberg blockade effects and the dynamics of Rydberg atoms. In this way, it is possible to obtain a variety of quantum phases, including vortex phases \cite{Self-generated_gauge_fields}, spin liquids \cite{Variational_approach_to_quantum_spin_liquids_rydberg,2D_quantum_chemistry,Emergent_gauge_theory_2023}, 
dimer and spin-ice models \cite{Quantum_spin_ice_dimer_models,Emerging_gauge_theories_2020} as well as to study phenomena in connection to many-body scar states and Hilbert space fragmentation \cite{emergent_gauge_theory_2024,Gauge_theory_description_rydberg,Emergent_symmetries_slow_QD_rydberg_chain_with_confinement}. 
Also, Rydberg atom platforms have been used to describe toy models in particle physics, such as Quantum Electro-Dynamics (QED) \cite{Confinement_of_quarks} and Quantum Cromo-Dynamics (QCD) \cite{QCD} in low dimensions, where non-perturbative effects such as confinement and string breaking can be analyzed. For example, the implementation of the one-dimensional Schwinger model \cite{Gauge_invariance_and_mass} has been analyzed theoretically in \cite{Lattice_gauge_string_dynamics_rydberg, Lattice_QED_rydberg_atoms}, realized in the seminal experiment described in \cite{Real-time_dynamics_LGT} or in the most recent works \cite{Probing_many_body_dynamics_51_atoms}, \cite{Observation_gauge_invariance_71-site} and \cite{String-breaking_mechanism_LSW}. 
It is also worth mentioning that non-abelian models have been taken into account, in \cite{QS_hadronic_states_rydberg} to describe hadronic states by means of Rydberg-dressed atoms and finally in
\cite{Hardware_efficient_QS_non-abelian_gauge_theories_qudits_rydberg} to implement a hardware-efficient quantum simulations via qudits on Rydberg platforms. 


\section{Chemistry applications}

\label{sec:Chemistry}

This epoch is characterized by a prominent advancement in the field of medicine, which is now able to devise and harness new tools, including High Performance Computers and machine learning techniques \cite{GML_medicine,GML_drug_discovery}, in order to achieve concrete steps forward in the creation and production of medicines and in the comprehension of the subtle mechanisms underlying the working processes of our organism. A complex task in this sense is drug discovery \cite{Overview_drug_discovery}, for which new approaches based on quantum computing are emerging, as described in section \ref{sec:Chemistry_molecular_docking},  together with molecular docking, necessary for drug discovery. Going to a more fundamental level of nature we find the field of quantum chemistry, which studies the main properties of molecules, including dynamics and energy spectrum \cite{Ab_initio_quantum_chemistry,Quantum_chemistry,2D_quantum_chemistry}, on which applications of Rydberg atoms are discussed in sec. \ref{sec:Chemistry_quantum_chemistry}.

\subsection{Molecular Docking and Drug Discovery}
\label{sec:Chemistry_molecular_docking}

The design of new drugs has always been a crucial task in modern medicine \cite{Overview_drug_discovery}. In recent years, new computational frameworks and techniques have been exploited to this aim, including machine learning \cite{GML_drug_discovery,Deep_learning_drug_discovery} and, ultimately, quantum computing \cite{Recent_advantages_QC_drug_discovery,QC_drug_discovery}. The steps of the process leading to the discovery of new drugs, as illustrated in \cite{Molecular_docking_review}, include the solution of a molecular docking problem \cite{Molecular_docking_old,Molecular_docking_review,Use_of_molecular_docking}. The latter consists in finding the best configuration of a molecule (the ligand) such that it perfectly ties with the binding site of a protein (the target) \cite{kirsopp2022quantum}. Each possible configuration found by the algorithm is evaluated with a score. To tackle a molecular docking problem on a neutral atoms device, it is required to map it into a Maximum Weighted Independent Set (MWIS), as explained in \cite{Molecular_docking}. In this paper, Garrigues et al. compared the performance of Variational Quantum Adiabatic Algorithm (VQAA), which makes use of a modified version of the Bayesian optimizer \cite{Bayesian_QAOA} called Hyperopt optimizer \cite{Hyperopt}, in order to optimize the hyperparameters (rise, sweep and fall time together with the detuning and the Rabi frequencies), with that of a supervised machine learning version of QAA, dubbed MLQAA, which predicts the best QAA parameters based on a training set of given VQAA runs. Each algorithm is evaluated by means of a scoring function, ranging between 0 and 1, which takes into account both energetic and geometric factors, evaluating how well the ligand fits the receptor's binding site and interaction energies between the ligand and the receptor. Because of the high variability of optimal QAA parameters, MLQAA average score is about one half of the one achieved by VQAA \cite{Molecular_docking}. Nevertheless, the score achieved by VQAA is almost 0.8 after 500 iterations and seems to be independent of the number of nodes (up to 11 at least). If this is going to be further confirmed, this approach may suggest potential scalability advantages on larger problem sizes. 

The prediction of water-solvent configuration is another fundamental issue in drug discovery, since water placement can drastically modify the binding capacity of the ligand and the thermodynamics of binding modes. Neutral atoms represent a great aid in this case, as shown in \cite{Drug_discovery}. The most probable configuration of water molecules in the ligand binding sites can be computed by mapping that problem into an Ising model, adopting the formalism of 3D Reference Interaction Site Model \cite{3D-RISM}. Subsequently, the Ising model can be solved with VQA on neutral atoms QPU, obtaining the desired solution.

\subsection{Quantum Chemistry}
\label{sec:Chemistry_quantum_chemistry}

Quantum chemistry \cite{Ab_initio_quantum_chemistry,Quantum_chemistry} tackles a wide variety of problems in a similar vast diversity of systems, including molecular dynamics, electronic structure of solids and molecular ground states preparation. In this regard, all quantum computing platforms emerged as a valuable tool to implement quantum chemistry simulations \cite{Trotter_quantum_chemistry,Kitaev_JW_transf_quantum_chemistry,Fault_tolerant_quantum_chemistry,Sublinear_scaling_quantum_chemistry,VQS_quantum_chemistry}, from ultracold atoms \cite{Quantum_chemistry_trapped_ion,seeking_QA_trapped_ion_quantum_chemistry, Analog_quantum_chemistry,Analogue_quantum_chemistry_simulation,Emulating_molecular_orbitals}, to superconducting qubits \cite{Variational_simulation_quantum_chemistry}, to NMR-based platforms \cite{Isomerization_quantum_chemistry}. 

A toy model of quantum chemistry is the hydrogen molecule. The electronic Hamiltonian, in second quantization, reads:
\begin{equation}
    H=\sum_{p,q}h_{pq}\hat{a}^{\dag}_p\hat{a}_q+\frac{1}{2}\sum_{p,q,r,s}h_{pqrs}\hat{a}^{\dag}_p\hat{a}^{\dag}_q\hat{a}_r\hat{a}_s
\end{equation}
where $\hat{a}_p^\dagger,{a}_p$ are creation/annihilation operators for electrons in the state index by $p$ and the coefficients $h_{pq}$ and $h_{pqrs}$ can be written in an integral form involving the corresponding wavefunctions \cite{hydrogen_molecule}. The coefficients can be computed numerically and, in addition, Jordan-Wigner transformations allow to rewrite the Hamiltonian in terms of Pauli strings (see eq. 22 in \cite{hydrogen_molecule} for the full Hamiltonian expression). Quantum algorithms can be therefore exploited for finding the ground state of the Hamiltonian, solving the problem. 

In general, the literature concerning quantum chemistry simulations is not as rich as one would expect, especially if compared to that regarding purely physical use cases and graph problems. Furthermore, Rydberg QPU has been specifically benchmarked as a candidate platform for molecular ground state preparation only in the last couple of years, with two different methods. \newline In the former approach \cite{Few-body_analog_QS_rydberg} Rydberg atoms play the role of electrons either of the $H_2$ molecule or of the He atom, the two systems taken into consideration, and the nuclear potential is simulated with a position-dependent AC Stark shift. This is what is called pseudo quantum chemistry, because, in the Rydberg atoms setting, the Coulomb interaction scales with a different power law with respect to the real one since it is implemented by leveraging Van der Waals interactions. By driving transitions between two hyperfine ground states to two Rydberg states, Malz and Cirac \cite{Few-body_analog_QS_rydberg} could extract the ground state energy of hydrogen and helium with electrons in singlet or triplet state. The ratio between the Rydberg decay rate and the hopping strength (i.e. the strength of the fermionic kinetic term) is the key factor determining how good the solutions will be. The increase of Rydberg state lifetime will therefore increase the performance of this algorithm.  \newline The latter approach \cite{Blueprint_VQE_rydberg}, in constrast, implements a VQE with Rydberg atoms in digital mode to compute the ground state energy of simple molecules, such as $LiH$ or $BeH_2$. The Hamiltonian of these molecules can be restated in terms of Pauli gates between a varying number of qubits, harnessing Jordan-Wigner or Bravyi-Kitaev transformations. Atomic positions on the register are classically optimized beforehand. A digital VQE with parameter splitting is performed and the energy is evaluated through a derandomization protocol \cite{Efficient_estimation_pauli_derandomization} based on the computation of local Pauli expectation values from an initial random set. The main drawbacks of this algorithm are the amount of time required to reach a small error ($\epsilon\leq 5\%$ in a day) and the scaling of the number of atoms on the molecular size. To this regard, in \cite{Blueprint_VQE_rydberg} the authors showed that for the relatively small molecules of water and methane containing 3 and 5 atoms, respectively, 14 and 18 qubits are required, and the Pauli terms in the Hamiltonian have a length of 595 and 1359, respectively. In the tested cases of $LiH$ and $BeH_2$,the number of qubits employed was 6 for both and the number of Pauli strings in the Hamiltonian was 118 and 165, respectively. Further studies are needed to investigate a possible quantum advantage brought by the mentioned algorithms.

\section{Hybrid Machine Learning  approaches}
\label{sec:machine_learning}

Machine Learning (ML) \cite{Machine_learning} is a flourishing and successful field of research since many decades \cite{Review_machine_learning}. In recent years, many research teams paved the way toward the enhancement of current machine learning capabilities by merging classical and quantum resources so that these two computing technologies compensate the weaknesses of each other \cite{Quantum_machine_learning,Challenges_in_QML}, entailing an overall improvement of the performance, at least for some specific problems \cite{QML_review_case_studies}. 

Graph Machine Learning (GML) algorithms \cite{GML_survey} are currently attracting a lot of attention due to the possible application on a wide range of problems of interest for society, including chemistry \cite{GML_chemistry} and medicine applications \cite{GML_medicine},  drug discovery \cite{GML_drug_discovery} and fingerprint recognition \cite{GML_fingerprint}. Rydberg-atoms-based technology has been harnessed specifically to enhance GML. In particular, QPU can be leveraged to extract quantum graph kernels \cite{QEK,Survey_graph_kernels}, devise new feature maps \cite{Quantum_feature_maps_GML} or compute encodings \cite{Enhancing_GNN_QC_encodings}. 

In \cite{QEK} an approach to graph kernels has been presented. It exploits a so called quantum graph, built up by embedding the data graph onto the atomic register, which is then evolved through the QAOA algorithms by a Hamiltonian, chosen between the Ising one and the XY one. At the end of the protocol, starting from the histogram of measurements, i.e. the probability distribution of the computational basis states, one is able to construct a quantum kernel function defined as the exponential of the Jensen-Shannon divergence of two probability distributions obtained for two different graphs, with which the classical model is subsequently fed and trained to solve the regression or the classification task of interest. Henry et al. \cite{QEK} proved that, on neutral atoms devices up to 12 nodes/qubits, the quantum-enhanced graph kernel, called Quantum Evolution Kernel (QEK) since it is derived from the evolution of a quantum system, outperforms purely classical kernel methods, such as Graphlet Sampling \cite{Graphlet_sampling_kernel} and Random Walk \cite{Random_walk_kernel},  on the several test datasets in terms of accuracy. In particular, on the fingerprint dataset only, the QEK has been tested even in the presence of a 5\% false positive and false negative rate \cite{QEK}) and it is still capable to outperform its competitors. Moreover, the emulated protocol outperforms classical kernels on all tested data sets up to 16 qubits \cite{QEK}.

These results have been expanded in \cite{Quantum_feature_maps_GML}. Employing a quantum evolution kernel defined as in \cite{QEK}, Albrecht et al. provided further proofs of the capability of graph kernels based on quantum computed observables by tackling a classification task concerning chemical compounds. The authors compared the QEK to the best state-of-the-art classical kernels, including the aforementioned Random Walk and Graphlet Sampling, but also the Shortest Path \cite{Shortest_path_kernel} and the SVM-$\theta$ kernel \cite{SVM-theta_kernel}, confirming that it was slightly better than SVM-$\theta$, the best classical kernel \cite{Quantum_feature_maps_GML}. 

Furthermore, the authors of \cite{Quantum_feature_maps_GML} proved that the QEK is capable of capturing geometrical features that classical kernels ignore they show  that their feature spaces are geometrically different. To practically show this, they devised the following graph data set: class A contains graphs whose nodes belong to a honeycomb lattice with inclusion of extra nodes with probability $p$. Class B graphs' nodes belong to the kagome lattice, including of extra nodes with probability $p$. As $p$ increases, classical kernels experiment some troubles in identifying the true class for graph data and they are outperformed by quantum counterparts. The more accurate way quantum kernels perceive data from a topological point of view can, in principle, lead to an advantage with respect to classical ones for graph classification and regression tasks. \newline
Following the footsteps of the previous references \cite{QEK, Quantum_feature_maps_GML}, Thabet et al. \cite{Enhancing_GNN_QC_encodings} devised a new quantum-computed encoding to be used on graph transformers model \cite{Graph_transformers,Graph_transformer_survey}, a newly-devised machine learning model based on neural network adapted for graph-structured data. By evolving, as in \cite{QEK}, the quantum graph with QAOA-like evolution operators, graph features can be recovered by computing spin-spin correlation matrices. QAOA parameters are trained together with the parameters of the classical model. The accuracy of graph transformers with quantum encodings turns out to be  on par with that obtained via classical encodings \cite{Enhancing_GNN_QC_encodings}. In particular, for the C-LADDER and S-PATTERN datasets, the accuracy of quantum encodings combined with graph convolutional network models \cite{GCN} is 100\%, outperforming classical encodings on the same model. 

An additional recent quantum machine learning application of neutral atoms concerns the financial sector \cite{Financial_risk}. With the aim of forecasting credit rating downgrades, a new quantum-enhanced classifier inspired by the QBoost algorithm \cite{qboost}, which leverages the solution of a QUBO problem \cite{QUBO}, has been compared to a classical random forest method \cite{Random_forest} based on random sampling with replacement \cite{Random_sampling_with_replacement} and simulated annealing \cite{simulated_annealing}. It turns out that the proposed algorithm was capable of outperforming random forest, reaching comparable precision and recall. The performance was similar to that of simulated annealing. 

Despite the improvement of ML algorithms with the aid of quantum processors is still at the very beginning due to the limited capacity of current quantum hardware, the employment of standard classical ML models working in synergy with quantum computers is really promising. Furthermore, the peculiarity of Rydberg arrays that allow for different desired atomic arrangement, has proved to be an essential benchmark for quantum-enhanced graph machine learning because of its capability to perceive geometrical graph features in a more sophisticated way, leading to better scores when tackling classification problems. In-progress hardware development under all points of view should enable, in the next future, the possibility of addressing problems with larger size and complexity, possibly unveiling the signs of quantum advantage.

\section{Conclusions} \label{se:conlusion}

Neutral atom devices, and quantum computers more broadly, represent powerful tools for the simulation of complex quantum systems, owing to their ability to directly manipulate and control quantum degrees of freedom. In contrast, classical computers rapidly become inefficient as system sizes grow, due to the exponential scaling of the underlying Hilbert space. Consequently, quantum processors are expected to play a pivotal role in extending our current computational capabilities and in addressing problems of increasing physical and mathematical complexity.

In this review, we have surveyed the rapidly evolving field of quantum computation and simulation based on neutral atom platforms, outlining their main applications across diverse scientific domains, including physics, chemistry, machine learning, and pharmacology. Beginning with an overview of the physical principles and architecture of neutral atom QPUs, we examined their benchmarking methodologies and performance metrics. Particular emphasis was placed on graph-based problems, especially the \textit{Maximum Independent Set} (MIS), which can be naturally encoded in neutral atom architectures and serves as a unifying framework for a broad range of optimization problems with significant industrial and informational relevance.

As highlighted throughout this work, further progress in hardware development remains essential. Key directions include achieving reliable local qubit addressability, improving the coherence and stability of Rydberg states, and scaling up register sizes to accommodate larger problem instances. Overcoming these challenges will be crucial to transitioning from proof-of-concept demonstrations to practical quantum advantage and to broadening the applicability of neutral atom quantum processors beyond current benchmark and emulation studies.

These systems can already provide concrete value for a class of near-term problems despite current hardware limitations, such as restricted coherence times and lack of local addressing. Medium-scale instances of Maximum Independent Set and related graph problems are native to analog encoding in the Rydberg Hamiltonian, as are QUBO models with moderate connectivity requirements. Other near-term applications include quantum-enhanced graph machine learning tasks, where geometric flexibility of atomic arrays enables expressive quantum feature maps.
\vspace{5pt}

\textit{This study was funded by the European Union - \textit{NextGenerationEU}, Mission 4, Component 2, Investment 1.4 in the framework of the CN HPC - \textit{“National Centre for HPC, Big Data and Quantum Computing”}, (CN HPC - CN00000013 – CUP D56G22000380006). The views and opinions expressed are solely those of the authors and do not necessarily reflect those of the European Union, nor can the European Union be held responsible for them. \\ E.E. acknowledges financial support from the MUR 2022-PRIN Project “Hybrid algorithms for quantum simulators”. E.E. and M.G. were partially funded by INFN IS-Quantum.}

\appendix
\section{Details on quantum computing with neutral atoms}
\label{Appendix:QC}
\subsection{Register loading and readout}
\label{Appendix:QC_register}
As briefly mentioned in the introduction of section \ref{sec:QC}, the quantum register is loaded in a vacuum chamber by trapping atoms with two counterpropagating lasers forming a so called \textit{optical tweezer} \cite{Optical_tweezers,Pasqal}. The tweezers can be then displaced in 2D or 3D arrays. Register size scaling depends only on the optical trapping power and performance. Hence, neutral atoms are considered a very promising platform for quantum computation scalable to thousands of logical qubits \cite{Quantum_computing_rydberg_atom_graphs}. \newline After the quantum processing phase, which is described in the next subsections and in which the optical tweezers are turned off to avoid any possible interference, the qubits are measured via fluorescence imaging. Basically, the tweezers are turned on, re-trapping the atoms in the ground state $|g\rangle$, while the excited atoms in $|r\rangle$ are anti-trapped and blown away. Therefore, the brightness (darkness) in the fluorescence image depends on the presence (absence) of the qubit in that register slot, hence on its quantum state. For hyperfine levels encoding, additional techniques involving transportation of an excited atom to a GS level and hyperfine selective detection are employed \cite{Parallel_low_loss_measurement,Fast_parallel_readout,Quantum_computing_rydberg_atom_graphs}.
\subsection{Digital computing mode}
\label{Appendix:QC_digital}
\begin{figure}[t]
    \centering
    \includegraphics[width=0.7\linewidth]{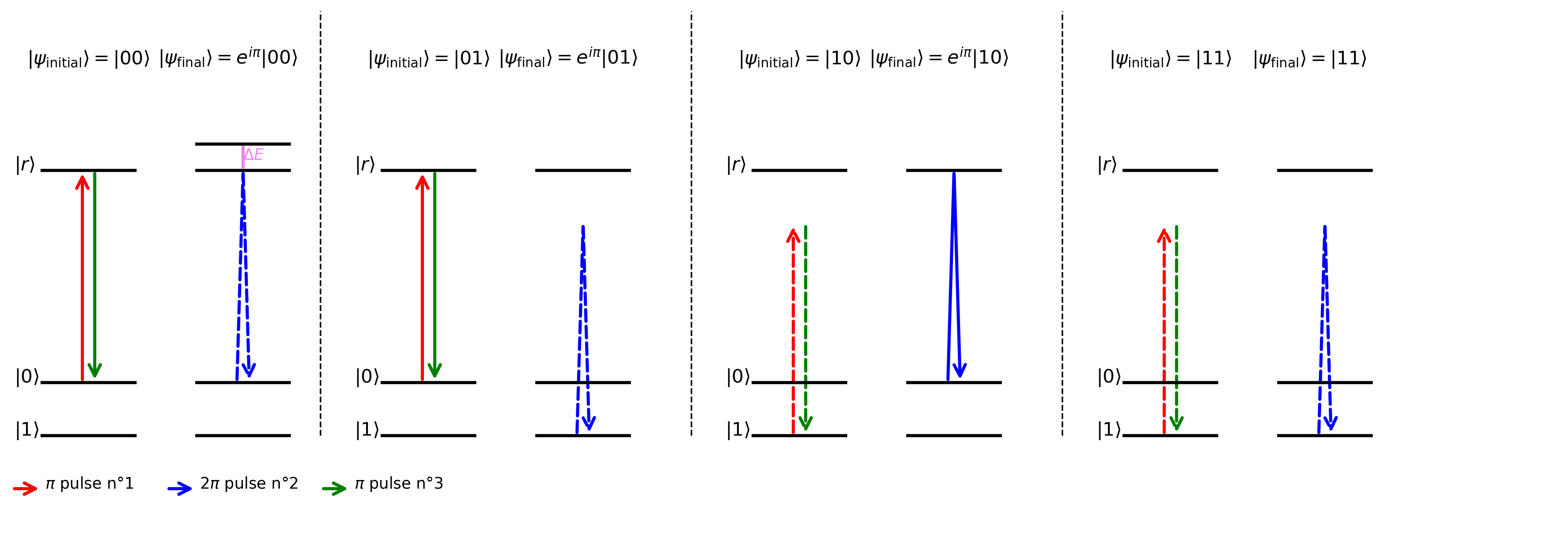}
    \caption{Pulse sequence required to implement CZ gate on neutral atoms in digital mode: a $\pi$ pulse is applied to the control qubit $q_c$, followed by a $2\pi$ pulse on the target qubit $q_t$ and another $\pi$ pulse to de-excite $q_c$. If the initial state is either $|00\rangle$, $|01\rangle$ or $|10\rangle$, the system acquires a phase $-1$ due to the successful application of the $2\pi$ pulse of step 2 or the combination of pulse 1 and 3. The system state is left invariant only when the initial state is $|11\rangle$ since all the pulses are off-resonant}
    \label{fig:CZ_pulses}
\end{figure}
The digital computing mode is characterized by hyperfine level encoding, harnessing the Rydberg state to entangle the qubits. It represents a universal computational framework, meaning that by applying quantum logical gates to the qubits, one can, in principle, reach any possible state in the Hilbert space $\mathcal{H}$ of the qubit states \cite{Nielsen_Chuang,Signal_processing,Digital_QS_rydberg_atoms,QI_rydberg,Quantum_networks}. In addition, every unitary transformation can be decomposed into single qubit rotations and CNOTs. \newline A laser pulse of duration $\tau$, with Rabi $\Omega$, detuning $\delta$ and phase $\phi$, will entail a local single qubit rotation about the $(x,y,z)$ axis of angles $(\Omega\tau\cos\phi,\Omega\tau\sin\phi,\delta\tau)$. \newline Controlled-NOT (or CX) can be implemented by applying two Hadamard gates interleaved by a CZ gate. The CZ is implemented with three consecutive laser pulses, two of them acting on the control and one on the target (see Fig. \ref{fig:CZ_pulses}). The idea is that when the starting state is either $|00\rangle,|01\rangle$ or $|10\rangle$ the whole two atoms system gets a phase equal to $e^{i\pi}=-1$ due to either the $2\pi$ pulse $2$ or the combination of $\pi$ pulses 1 and 3. Notice that if the initial state is $|00\rangle$, pulse 2 becomes off-resonant due to the shift of the Rydberg level on the target qubit. In the $|11\rangle$ case, all the pulses are off-resonant and no phase is acquired. Therefore, this protocol allows to apply the unitary transformation $diag(-1,-1,-1,1)=e^{i\pi}CZ$ where the global phase $e^{i\pi}$ can be neglected.  
\subsection{Analog quantum computing}
\label{Appendix:QC_analog}
The analog mode is characterized by the encoding of the basis state into the so called "Ground-Rydberg" basis, that is, the $|0\rangle$ state is the ground state of the rubidium atom while the $|1\rangle$ state is encoded in the excited Rydberg state. Moreover, the state of the system is evolved through a global Hamiltonian whose form is:
\begin{equation}
    H=\frac{\Omega}{2}\sum_i(\cos(\phi)\sigma^x_i-\sin(\phi)\sigma^y_i)-\frac{\delta}{2}\sum_i\sigma^z_i+\sum_{\langle ij\rangle}U_{ij}n_in_j
    \label{eq:Hamiltonian_pasqal}
\end{equation}
being $\phi$ the relative phase, $U_{ij}=\frac{C_6}{r_{ij}^{6}}$ the interaction coefficient between atoms and $n_i=\frac{1+\sigma^z_i}{2}$ the occupation number operator, having set $\hbar=1$. The coefficient $C_6=5,420,503\: \frac{{\mu m}^6\dot rad}{\mu s}$ for two atoms in the $|70S_{1/2}\rangle$ state \cite{Aquila_Quera}. \newline In this setup, quantum gates cannot be performed since the Rabi and the detuning frequencies are global and the laser acts the same way on all the atomic register. \newline Another implementation of a global Hamiltonian in analog mode involves a third possible choice for the basis encoding. This is called the "XY" basis, and the qubit basis states are encoded into two Rydberg levels which are dipole-coupled, i.e. experiencing dipole-dipole interactions, such as $|nS\rangle$ and $|nP\rangle$, thus having a angular momentum difference $\delta l=1$ \cite{Pasqal,QS_QC_rydberg_interacting_qubits}. In this setup, the driving Hamiltonian reads:
\begin{equation}
    H=\frac{\Omega}{2}\sum_i(\cos(\phi)\sigma^x_i-\sin(\phi)\sigma^y_i)-\frac{\delta}{2}\sum_i\sigma^z_i+2\sum_{\langle ij\rangle}\frac{C_3}{r_{ij}^3}(\sigma^x_i\sigma^x_j+\sigma^y_i\sigma^y_j)
    \label{eq:Pasqal_XY}
\end{equation}
where now the interaction coefficient $C_3=\frac{3700 {\mu m}^3}{\mu s}$ for the choice of Rydberg encoding levels $|0\rangle=|62D_{3/2}, m_j=3/2\rangle$ and $|1\rangle=|63P_{1/2},m_j=1/2\rangle$. The Hamiltonians in eqs. (\ref{eq:Hamiltonian_pasqal}) and (\ref{eq:Pasqal_XY}) are well-suited to study quantum many body theory models including the well-known Ising model \cite{Ising_model} and the Heisenberg model \cite{Heisenberg_model}. 
\subsection{A glimpse on partial addressing}
\label{Appendix:QC_partial}
Although the final goal of quantum hardware improvement is the implementation of digital, quantum error corrected, gates, which would allow a universal quantum computational framework and the possibility to compare quantum processors with classical ones aiming at an effective quantum advantage, if not supremacy, an acceptable near-term midway compromise consists in the so called \textit{partial addressability}, which should become available for the implementation on real quantum device in a very short time. It leverages the possibility to locally tune the detuning frequency for each atom site, hence $\delta\rightarrow\delta_i$. This would unlock the possibility to simulate a wider class of models whose magnetic properties depend on local fields, for instance the preparation of the GHZ state on a lattice with open boundary conditions with high fidelity \cite{Generation_schrodinger_cat_state_rydberg}, or amorphous magnets \cite{Amorphous_quantum_magnets}, together with a plethora of other models that are discussed in the main text of this review.
\subsection{Noise sources in neutral atoms devices}
\label{Appendix:QC_noise}
\begin{figure}[t]
    \centering
    \includegraphics[width=0.7\linewidth]{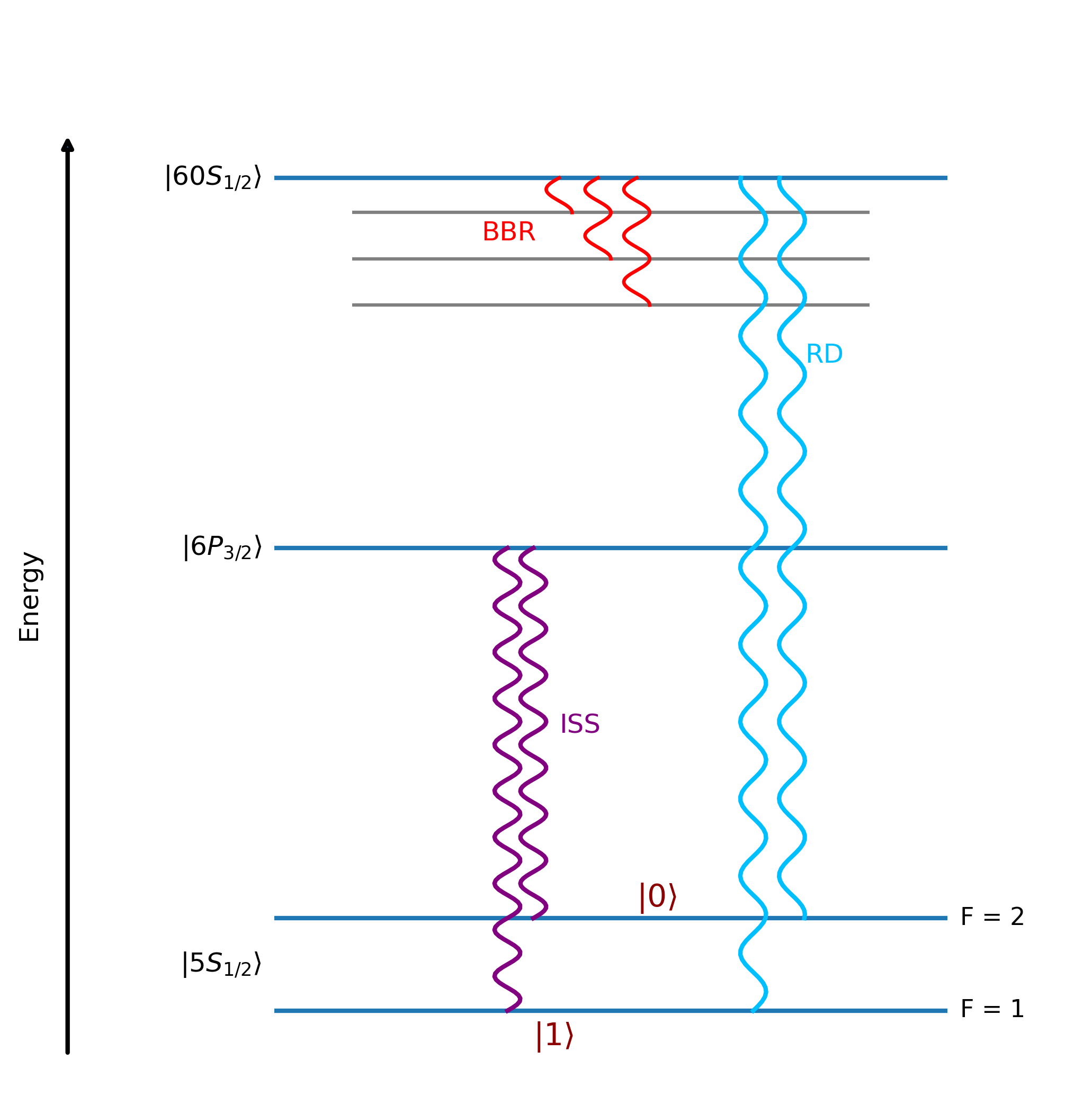}
    \caption{Main noise sources in neutral atom devices: Black-Body Radiation (BBR, red) consists in the excitation of the system from the Rydberg state $|nS\rangle$ used for computation to a Rydberg $|nP\rangle$ state; Radiative Decay (RD, light blue) consists in the de-excitation of the atom from the Rydberg state to the ground state manifold; Intermediate State Scattering (ISS, purple) consists in the de-excitation of the atom from the intermediate state to the ground state manifold}
    \label{fig:noise}
\end{figure}
The main noise sources are shown in Fig. \ref{fig:noise}. \newline The primary source of errors is due to \textit{Black-Body Radiation} (BBR) (red Fig. \ref{fig:noise}). When this occurs, an excited atom "jumps" to a nearby Rydberg level compatibly with selection rules. Since this level cannot be de-excited to $|0\rangle$ to which it is not coupled and cannot decay to the ground state within the time of applications of subsequent gates, this entails a continuous blockade effect on nearby atoms, potentially resulting in correlated $Z$-type errors. The overall effect of BBR-induced errors can be suppressed by increasing the principal quantum number $n$ of Rydberg levels or decreasing the temperature since the rate of BBR transition is proportional to $T$ and inversely proportional to $n^2$ \cite{Hardware_efficient_QC_rydberg}. \newline When a spontaneous emission event occurs in an excited atoms, we talk about \textit{Radiative Decay} (RD). The atom in the Rydberg state decays into one of the levels in the ground state manifold (light blue Fig. \ref{fig:noise}). This entails $Z$-type correlated errors if the RD event occur during entangling gates. Furthermore, RD rate is temperature-independent and decreases with $n$. Hence, RD dominates over BBR in the low $T$ and small $n$ limit \cite{Hardware_efficient_QC_rydberg}. The RD error class includes a peculiar subclass containing the so called \textit{Intermediate state scattering}, consisting in decays from intermediate state, e.g. $6P_{3/2}$ (purple, Fig. \ref{fig:noise}) to the ground state manifold. This error can be suppressed by increasing laser detuning and laser power \cite{Hardware_efficient_QC_rydberg}. \newline In conjunction with the noise sources directly correlated with the physical hardware, experimental imperfections can also provoke an additional plethora of errors that we will briefly list here. \textit{Fluctuations of laser intensity, frequency and phase} and \textit{atom loss} are the most significant errors. Nevertheless, they can be modeled using $Z$ operators hence addressed with simple QEC codes together with BBR and RD. Finite atomic temperature (which causes Doppler shifts) and \textit{temperature-induced errors} can be suppressed by increasing the interaction strength between atoms. Leakage to other hyperfine ground states, called \textit{Raman transitions} can be suppressed by detuning the optical tweezers light. $X$ and $Y$ error types can occur due to miscalibrations of the laser but they can be removed with proper pulse composition. In the next subsections we will see how to address errors from both hardware and software points of view.
\section{Quantum Optimization Algorithms}
\label{appendix:Algo}
In this appendix we will present the three pillars of quantum optimization algorithms, i.e. the \textit{Quantum Approximate Optimization Algorithm} (QAOA) \cite{farhi_qaoa}, the \textit{Quantum Adiabatic Algorithm} (QAA) \cite{QAA,QAA_rydberg_dressed} and the \textit{Variational Quantum Eigensolver} \cite{VQE_peruzzo,VQE}, with an eye looking at possible applications on neutral atoms \cite{VQA_pulse_opt_rydberg}. Afterwards, we will analyze further developments introduced by newly-devised approaches toward the optimization of register mapping techniques to embed the atoms in a convenient geometric configuration based on the problem we aim to solve. 
\subsection{The Quantum Approximate Optimization Algorithm}
\begin{figure}[t]
    \centering
    \includegraphics[width=0.7\linewidth]{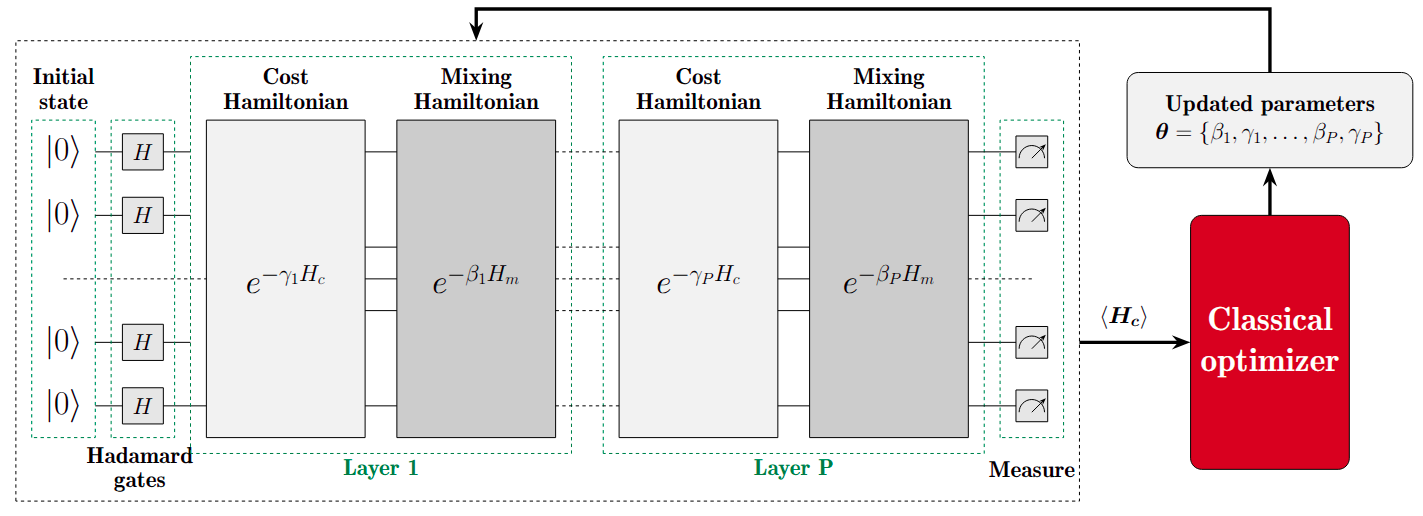}
    \caption{Scheme of the Quantum Approximate Optimization Algorithm. The system is prepared in a superposition state via Hadamard gates and alternately evolved with the cost Hamiltonian of the problem $H_C$ and a mixing Hamiltonian $H_M$ which is chosen such that $[H_C,H_M]\neq 0$. The evolution steps are parameterized by angles $\gamma_1,...,\gamma_p$ (for cost layers) and $\beta_1,...,\beta_p$ (for mixing layers). At each iteration the energy of the system is computed and a classical optimizer minimizes the energy by updating the parameter set up to convergence}
    \label{fig:qaoa}
\end{figure}
The Quantum Approximate Optimization Algorithm, a.k.a. QAOA, has been introduced by Farhi et al. in 2014 \cite{farhi_qaoa}. Fig. \ref{fig:qaoa} provides a scheme of its workflow. The core idea is to drive the system evolution with two non-commuting Hamiltonians, called the \textit{cost Hamiltonian} $H_C$, i.e. the energy operator whose ground state encodes the solution of the problem, and the \textit{mixer Hamiltonian} $H_M$. Since an overwhelming majority of the problem Hamiltonians include terms which consist of Pauli $\sigma^z$ operators, the typical choice for the mixer is $H_M=\sum_i\sigma^x_i$, because it does not commute with $\sigma^z$, i.e. $[\sigma^x_i,\sigma^z_j]\neq 0$. If the cost Hamiltonian itself consists of two non-commuting terms, the evolution operators to be applied will be driven by these two terms. This is the case when one tries to find the ground state of physical models like the Ising model in transverse field. The QAOA imposes to set the initial state of the system to the balanced superposition of all basis states $|+\rangle^{\otimes N}$ where $|+\rangle=\frac{|0\rangle+|1\rangle}{\sqrt{2}}$ is the eigenstate of $\sigma^x$ with eigenvalue 1. Parameterized evolution operators driven by the cost and the mixer Hamiltonians are then applied in alternated fashion for $p$ times, where $p$ is called the "depth" of the circuit, a.k.a. its number of layers. At the end of the circuit the energy, i.e. the cost value, is estimated from repeated measurements as $\frac{1}{N}\sum_i\langle\psi_i|H_C|\psi_i\rangle$, where $|\psi_i\rangle$ is the outcome bitstring of the $i$-th measure and $N$ is the number of measurements. The parameters are then updated by a classical optimizer and the procedure is repeated until the optimizer converges to the best circuit parameters. \newline In neutral atoms devices, the cost Hamiltonian for typical problems is the interaction term of eq. (\ref{eq:Hamiltonian_pasqal}) while the mixer term is the $\Omega$-driven one in the same eq. Notably, since one cannot turn off the interaction during the computation because of the static atomic positions in the register, the QAOA can be implemented at the price of some imperfections. Despite that, since QAOA is a native algorithm for current analog mode, it is nonetheless still considered a promising asset and a benchmark for neutral atom-based quantum computing architectures \cite{Multi-qubit_entanglement,Fahri_QAOA_supremacy}. \newline A recent improvement for QAOA on neutral atoms has been proposed in \cite{Quantum_opt_four_body_rydberg_gates}, where optimization problems whose Hamiltonians contain interaction terms involving more than 2 qubits can be also tackled with a proper mapping called "LHZ parity encoding" \cite{QAA_LHZ}. It consists of substituting the many-body interaction term $J_{ijk...}\sigma^z_i\sigma^z_j\sigma^z_k...$ with a single-qubit term $J_{\mu}\sigma^z_{\mu}$, hence remapping the long-range interactions into local fields. The price to pay is an additional "plaquette" term in the QAOA Hamiltonian which is implemented via adiabatic protocol. The key idea is to overcome the constraint of local interactions dictated by the architecture, widening the spectrum of optimization problems addressable with QAOA.
\subsection{The Quantum Adiabatic Algorithm}
The Quantum Adiabatic Algorithm \cite{QAA} leverages the quantum adiabatic theorem \cite{Quantum_adiabatic_theorem_proof} to solve optimization problems encoded into the ground state of a problem Hamiltonian $H$. In a nutshell, the quantum adiabatic theorem states that: \newline\newline \textit{Given a time-dependent Hamiltonian $H(s)$ where $s\in[0,1]$ and $\psi(s)$ is its ground state, if, for any $s$, the energy spectrum is gapped, an adiabatic evolution driven by $H(s)$ on $\psi(s)$ will result in $\psi(1)$, i.e. the ground state of the final Hamiltonian.} \newline\newline A quantum system can therefore be driven from an easy-to-prepare ground state of an initial Hamiltonian, call it $H_I$, to the ground state of the problem Hamiltonian, encoding the solution of the problem, call it $H_P$, provided that the energy levels of the Hamiltonians do not intertwine. The evolution follows the equation:
\begin{equation}
    H(s)=H_P(s)+H_I(1-s).
\end{equation}
Ref. \cite{Quantum_adiabatic_theorem_proof} provides a quantitative proof of the theorem whereas \cite{QAA} provides a complete review of how to implement quantum algorithms, including Grover's \cite{Grover_algo}, Deutsch-Jozsa (DJ) \cite{Deutsch-Jozsa_algorithm,Deutsch-Jozsa_algo2} and the Bernstein-Vazirani (BV) algorithm \cite{Bernstein-Vazirani_algo} in an adiabatic version. \newline An initial step toward the implementation of QAA on neutral atoms is dated back to 2013 \cite{QAA_rydberg_dressed}. A more sophisticated version of QAA which leverages two atomic species, i.e. rubidium and caesium, together with the LHZ parity encoding \cite{QAA_LHZ} mentioned in the previous section, is discussed in \cite{Coherent_annealer_rydberg}.
\subsection{The Variational Quantum Eigensolver}
The Variational Quantum Eigensolver has been devised by Peruzzo et al. in 2014 \cite{VQE_peruzzo,Photonics}. The goal is, once again, to find the ground state of a given Hamiltonian $H$. The algorithm harnesses a parameterized quantum circuit from which one can extract a quantum state on which to compute the expectation value of $H$, i.e. the energy. A classical optimizer is exploited for the minimization of the energy. Unlike the QAOA, the circuit overall structure is not established a priori but one can, in principle, devise numerous ansatzes, which may achieve a different accuracy. VQE finds a lot of applications in physics and chemistry, well summarized in \cite{VQE} together with error mitigation techniques. Later on in this review we will encounter the VQE again for quantum chemistry benchmarks on Rydberg atoms \cite{Blueprint_VQE_rydberg}.
\bibliographystyle{unsrt}
\bibliography{referenze}
\nocite{*}




\end{document}